\documentclass[twoside,twocolumn,9pt]{article}
\usepackage{extsizes}
\usepackage[super,sort&compress,comma]{natbib} 
\usepackage[version=3]{mhchem}
\usepackage[left=1.5cm, right=1.5cm, top=1.785cm, bottom=2.0cm]{geometry}
\usepackage{balance}
\usepackage{times,mathptmx}
\usepackage{sectsty}
\usepackage{graphicx} 
\usepackage{lastpage}
\usepackage[format=plain,justification=justified,singlelinecheck=false,font={stretch=1.125,small,sf},labelfont=bf,labelsep=space]{caption}
\usepackage{float}
\usepackage{fancyhdr}\usepackage{epsfig,graphics,amssymb,amsmath,subeqnarray,amsthm,latexsym}

\usepackage{fnpos}
\usepackage[english]{babel}
\addto{\captionsenglish}{%
  
}
\usepackage{array}
\usepackage{droidsans}
\usepackage{charter}
\usepackage[T1]{fontenc}
\usepackage[usenames,dvipsnames]{xcolor}
\usepackage{setspace}
\usepackage[compact]{titlesec}
\usepackage{hyperref}

\newcommand{\mean}[1]{\left\langle{#1}\right\rangle}
\usepackage{amssymb}

\usepackage{tikz}
\usetikzlibrary{arrows, backgrounds, shapes, positioning}
\tikzset{
    >=stealth',
    bubble/.style={
           ellipse,
           draw=black, very thick,
           text centered},
    pil/.style={
           ->,
           thick,
           shorten <=2pt,
           shorten >=2pt,}
}

\definecolor{cream}{RGB}{222,217,201}

\begin{document}

\pagestyle{fancy}
\thispagestyle{plain}
\fancypagestyle{plain}{

\fancyhead[C]{\flushright Accepted for publication in {\it Soft Matter}, 2019}
\renewcommand{\headrulewidth}{0pt}
}

\makeFNbottom
\makeatletter
\renewcommand\LARGE{\@setfontsize\LARGE{15pt}{17}}
\renewcommand\Large{\@setfontsize\Large{12pt}{14}}
\renewcommand\large{\@setfontsize\large{10pt}{12}}
\renewcommand\footnotesize{\@setfontsize\footnotesize{7pt}{10}}
\makeatother
\def\bka{{\boldsymbol \kappa}}
\renewcommand{\thefootnote}{\fnsymbol{footnote}}
\renewcommand\footnoterule{\vspace*{1pt}%
\color{cream}\hrule width 3.5in height 0.4pt \color{black}\vspace*{5pt}} 
\setcounter{secnumdepth}{5}

\makeatletter 
\renewcommand\@biblabel[1]{#1}            
\renewcommand\@makefntext[1]%
{\noindent\makebox[0pt][r]{\@thefnmark\,}#1}
\makeatother 
\renewcommand{\figurename}{\small{Fig.}~}
\sectionfont{\sffamily\Large}
\subsectionfont{\normalsize}
\subsubsectionfont{\bf}
\setstretch{1.125} 
\setlength{\skip\footins}{0.8cm}
\setlength{\footnotesep}{0.25cm}
\setlength{\jot}{10pt}
\titlespacing*{\section}{0pt}{4pt}{4pt}
\titlespacing*{\subsection}{0pt}{15pt}{1pt}

\fancyfoot{}
\fancyfoot[RO]{\footnotesize{\sffamily{1--\pageref{LastPage} ~\textbar  \hspace{2pt}\thepage}}}
\fancyfoot[LE]{\footnotesize{\sffamily{\thepage~\textbar\hspace{3.45cm} 1--\pageref{LastPage}}}}
\fancyhead{}
\renewcommand{\headrulewidth}{0pt} 
\renewcommand{\footrulewidth}{0pt}
\setlength{\arrayrulewidth}{1pt}
\setlength{\columnsep}{6.5mm}
\setlength\bibsep{1pt}

\makeatletter 
\newlength{\figrulesep} 
\setlength{\figrulesep}{0.5\textfloatsep} 

\newcommand{\topfigrule}{\vspace*{-1pt}%
\noindent{\color{cream}\rule[-\figrulesep]{\columnwidth}{1.5pt}} }

\newcommand{\botfigrule}{\vspace*{-2pt}%
\noindent{\color{cream}\rule[\figrulesep]{\columnwidth}{1.5pt}} }

\newcommand{\dblfigrule}{\vspace*{-1pt}%
\noindent{\color{cream}\rule[-\figrulesep]{\textwidth}{1.5pt}} }

\makeatother

\twocolumn[
  \begin{@twocolumnfalse}
\vspace{3cm}
\sffamily
\begin{tabular}{m{4.5cm} p{13.5cm} }

& 
\noindent\Large{\textbf{A stochastic model for bacteria-driven micro-swimmers}} \\
\vspace{0.3cm} & \vspace{0.3cm} \\

 & \noindent\large{Christian Esparza L\'opez, Albane Th\'ery and Eric Lauga\textit{$^{\ddag}$}} \\

& \noindent\normalsize{

 Experiments have recently shown the feasibility of utilising bacteria as micro-scale robotic devices, with  special attention paid to the development of bacteria-driven micro-swimmers taking   advantage of  built-in actuation and
 sensing mechanisms of  cells.   Here we propose a stochastic fluid dynamic model to describe analytically and computationally  the dynamics of microscopic particles driven by the motion of surface-attached  bacteria  undergoing
 run-and-tumble motion. We compute analytical expressions for the rotational diffusion coefficient, the swimming speed and the effective diffusion coefficient. At short times, the mean squared displacement (MSD) is proportional to
 the square of the swimming speed, which is independent of the particle size (for fixed density of attached bacteria) and scales linearly with the number of attached bacteria; in contrast, at long times the MSD scales quadratically
 with the size of the swimmer and is independent of the number of bacteria. We then extend our result to the situation where the surface-attached bacteria undergo chemotaxis within the linear response regime. We demonstrate that  
 bacteria-driven particles are capable of performing artificial chemotaxis, with a chemotactic drift velocity linear in the chemical concentration gradient and independent of the size of the particle. Our results are validated
 against numerical simulations in the Brownian dynamics limit and will be relevant to the optimal design of micro-swimmers for biomedical applications.

} \\

\end{tabular}

 \end{@twocolumnfalse} \vspace{0.6cm}

  ]
 \renewcommand*\rmdefault{bch}\normalfont\upshape
\rmfamily
\section*{}
\vspace{-1cm}

\footnotetext{\textit{Department of Applied Mathematics and Theoretical Physics, University of Cambridge, Cambridge CB3 0WA, United Kingdom.}}
 \footnotetext{\ddag~e.lauga@damtp.cam.ac.uk}

\section{\label{sec:intro}Introduction}
\begin{figure*}
   \centering
   \includegraphics[width=0.7\textwidth]{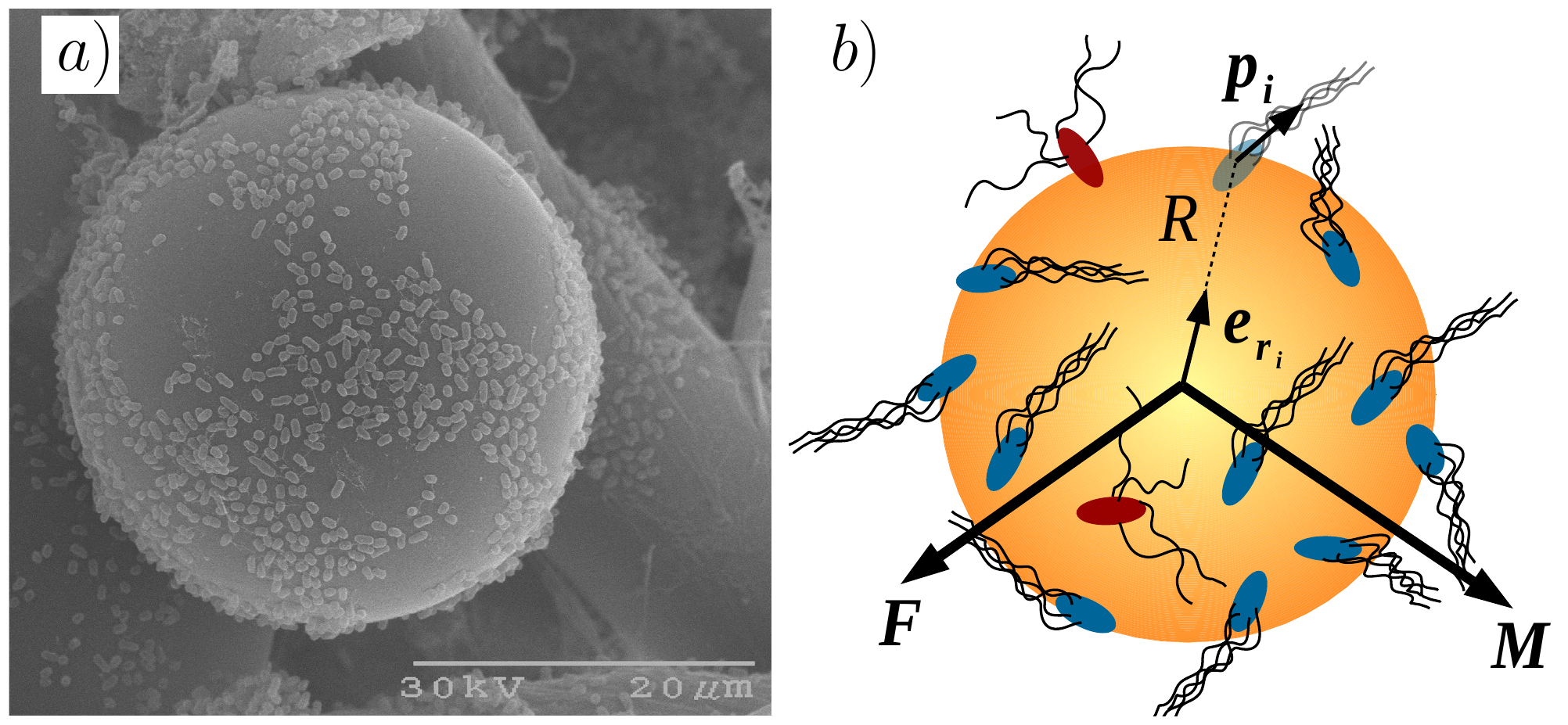}
   \caption{Bacteria-driven micro-swimmer. (\textit{a}) Scanning electron microscope (SEM) image of a $30\, \mu$m diameter bead with surface-attached bacteria. The individual cells are seen as small dots on the smooth surface of the
   bead (reprinted by permission from Kim et al.\cite{Kim2012}. Copyright 2012 from Springer Nature). (\textit{b}) Schematic representation of a bacteria-driven particle of radius $R$. The unit vectors $\mathbf{e}_{r_i}$ and $\mathbf{p_i}$ define the position and orientation of the $i$-th bacterium. The
   total hydrodynamic force, $\mathbf{F}$, and   torque, $\mathbf{M}$, are the sum of the applied forces and torques by each bacterium, $-f\mathbf{p}_i$ and $-Rf\left(\mathbf{e}_{r_i}\times\mathbf{p}_i\right)$ 
   respectively.}
   \label{fig:micro_swimmer}
  \end{figure*}

  Miniaturisation of actuators and efficient power sources are two of the biggest technical challenges in the design and fabrication of microscopic  robots \cite{Sitti2009, Martel2012}.  As is often the case, Nature can offer insight
  into overcoming these challenges. Flagellated bacteria,   such as the well-studied \textit{E.~coli}, are known to be   efficient swimmers with intricate sensing capabilities \cite{Berg2004} and they have  inspired scientists to
  integrate living cells and synthetic components into bio-hybrid devices \cite{HOSSEINIDOUST201627, Carlsen2014, Ceylan2017, Julius2009, Darnton2004, Martel2012, Katuri2017, Schwarz2017, Wang2018, Bastos2018}. Bacteria-driven micro-swimmers have received special attention due to
  their potential applications in medicine such as targeted drug delivery \cite{zhuang2016, arabagi2011modeling, zhuang2017propulsion, Behkam2008, Darnton2004, Schauer2018, Cho2012, NGUYEN2016217}.  Commonly, a micro-swimmer consists of a synthetic substrate
  such as a micro-plate \cite{Julius2009, Steager2011} or a micro-bead \cite{Darnton2004, Behkam2008, Behkam2007, Schauer2018} with just a few or many  bacteria attached to their surfaces (see experimental example in
  Fig.~\ref{fig:micro_swimmer}a). The attachment of the cells to the surface is often random but can be controlled by patterning the   substrate using chemical  or physical techniques \cite{Carlsen2014, Morgan2015}. The dynamics of a
  passive bead driven by surface-attached bacteria is a result of  the precise   behaviour of each cell. When free swimming, a peritrichous  bacterium such as \textit{E.~coli} moves in a series of nearly straight paths (`runs'),
  interrupted by quick changes in orientation (`tumbles'). This so-called run-and-tumble motion consists therefore of two stages: (i) the running phase, in which all the helical flagellar filaments rotate counterclockwise (when
  measured from behind the cell) and form a bundle which aligns with the cell body and whose rotation in the fluid propels the cell forward; (ii) the tumble phase, in which at least one flagellar filament rotates in the  clockwise
  direction and leaves the  bundle resulting in a reorientation of the whole cell \cite{Berg2004, Lauga2016}. During the running phase, the bacterium experiences a constant propulsive force and torque along the same direction.
  Both, force and torque, balance with the corresponding hydrodynamic translational and rotational drags, so that the cell is overall force- and torque-free. In contrast, in the tumbling phase the bacterium experiences no propulsive
  force but a torque which reorients the cell body \cite{zhuang2017propulsion}.
  The running and tumbling events for \textit{E.~coli}  are known to be Poisson distributed \cite{BLOCK1982215} with each run lasting for approximately one second while tumbles last just a tenth of a second \cite{Lauga2016}. These
  times can be altered by chemical concentration gradients  in the environment surrounding the cell. Bacteria such as \textit{E.~coli} can measure temporal differences in concentration of certain chemicals during their locomotion,
  for example  aspartate, and reduce their tumbling rate if moving up the gradient \cite{Segall1986, Schnitzer1993}.  As a result, the trajectory of a swimming bacterium is an isotropic random walk in homogeneous environments and a
  biased random walk when in the presence of a chemical gradient (if the cell is sensitive to it). The bias in the direction of motion is known as chemotaxis and when displayed by bacteria attached to passive particles it can
  potentially provide chemotactic abilities to the particles themselves \cite{zhuang2016, Park2013}. Indeed, there are many other taxis techniques which can be used to control the trajectories of synthetic micro-swimmers, such as
  phototaxis \cite{Steager2007} and magnetotaxis \cite{Carlsen2014_1}. However, for medical applications chemotaxis appears to be the most natural, as not only it does not require the use of external force fields but can also be
  linked to the ubiquitous presence of chemical gradients in the human body.
 
  Helical trajectories of particles with surface-attached bacteria have been observed in experiments with small beads on which only one or two  bacteria can be attached \cite{zhuang2017propulsion, Edwards2013}.  This suggests that
  the cells, despite being attached to a surface, still follow a run-and-tumble dynamics, and each bacterium applies a constant force and constant torque to the bead. Following this observation,  a number of mathematical models have
  been proposed to describe the motion of bacteria-driven micro-swimmers \cite{Julius2009, arabagi2011modeling, zhuang2017propulsion}. The  swimmers are modelled as spherical particles actuated by random forces and torques taken as
  the sums of the individual propulsive  forces and torques applied by each bacterium. Given the small velocity scales, the motion occurs at low Reynolds number and is governed by Stokes laws which relate linearly the applied forces
  and torques to the linear and angular velocities of the bead. Although these models have been validated numerically against experimental results \cite{zhuang2016, arabagi2011modeling, zhuang2017propulsion, Behkam2008, Darnton2004},
  very few analytical expressions have been derived. For example, it is known that for large number of surface-attached bacteria $N$, the swimming speed increases as $\sqrt{N}$ \cite{Behkam2008, arabagi2011modeling} but the dependence
  on the size of the micro-swimmer is unclear.

  A detailed mathematical analysis of the motility properties of the micro-swimmers is essential for their optimal design and fabrication. In this study, we thus develop  a stochastic  model for bio-hybrid spherical micro-swimmers with
  \textit{E.~coli} bacteria   attached to their surface.  We derive analytical expressions for the rotational diffusion coefficient  and the mean squared displacement (MSD), based on the following assumptions: (i) low-Reynolds number
  flow, (ii) a large number of uniformly attached bacteria, (iii) negligible thermal noise compared to the random activity of each cell and (iv) run-and-tumble dynamics for each bacterium. We first consider the situation where the
  chemical environment is homogenous and next investigate the effects of external chemical gradients, assuming a linear response of each bacterium. In particular, we show rigorously that the individual capabilities of \textit{E.~coli}
  cells to move up   concentration gradients are inherited by the micro-swimmers on which they are attached. Our analytical results are validated against numerical simulations (using a standard Brownian dynamics framework) and past
  experimental results  \cite{zhuang2016, arabagi2011modeling, zhuang2017propulsion, Behkam2008, Darnton2004}.
 \section{\label{sec:numerical} Mathematical model and numerical simulations}
  \subsection{\label{subsec:model} Micro-swimmers in homogeneous environments}
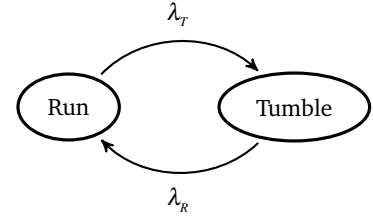
\begin{figure}
    \centering
    \tikzstyle{elp}=[ellipse, draw, thick]
    \begin{tikzpicture}[inner sep=2mm]
     \node [bubble] (run) {Run};
     \node (dummy1) [right of=run] {};
     \node (dummy2) [right of=dummy1] {};
     \node [bubble] (tumble) [right of=dummy2] {Tumble}
      edge [pil, <-, bend right=45] node [auto, swap] {$\lambda_{_T}$} (run)
      edge [pil, ->, bend left=45] node [auto] {$\lambda_{_R}$} (run); 
    \end{tikzpicture}
    \caption{The run-and-tumble motility pattern as a two-state Markov chain with transition rates
    $\lambda_{_R}$ and $\lambda_{_T}$   from the tumbling to the running state and
    vice-versa, respectively. Typical values of these rates for \textit{E.~coli} in homogeneous environments (no chemical gradients) are
    $\lambda_{_T}^{-1}=0.9~{\rm s}$ and $\lambda_{_R}^{-1}=0.1 ~{\rm s}$ \cite{BLOCK1982215}.}
    \label{fig:r-n-t}
   \end{figure}

   Following previous studies \cite{zhuang2017propulsion, arabagi2011modeling}, we model the micro-swimmers as passive spherical particles of radius $R$ with a number $N$ of uniformly distributed bacteria attached to their surface.
   Reported densities of attachment in experiments range from one bacterium per $12 \,\mu \text{m}^2$ up to one per $7 \,\mu \text{m}^2$ \cite{Behkam2008}. The bacteria are assumed to be fixed in position and orientation with respect
   to the surface of the particle. This condition is satisfied in practice by using strong chemical binding such as streptavidin-biotin interactions \cite{zhuang2017propulsion}. Previous numerical investigations suggest
   that the direction of the flagellar bundle is unaffected by the fluid flow in the vicinity of the swimmer \cite{arabagi2011modeling}, therefore the flagellar bundle is assumed to align with the orientation of the cell body with no
   change as the swimmer moves. 
   \begin{figure}
    \centering
    \includegraphics[width=0.8\columnwidth]{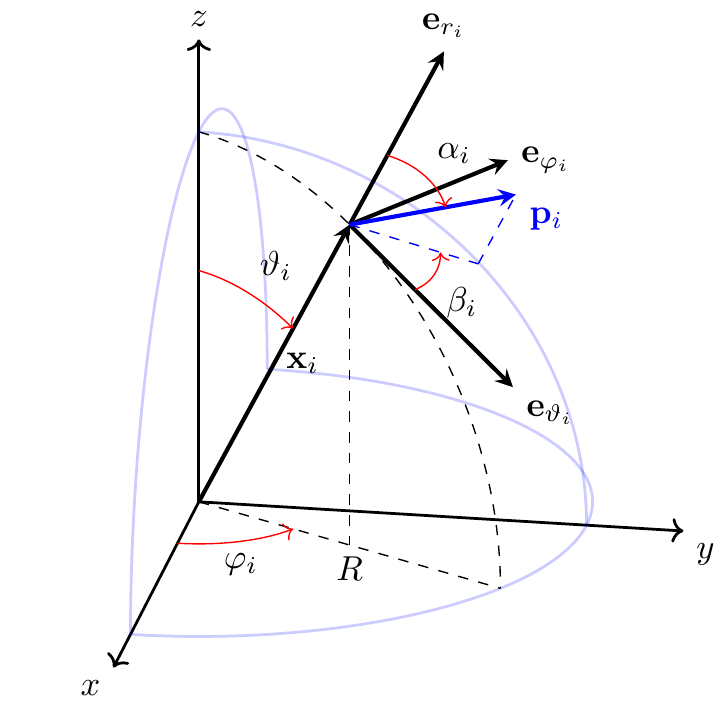}
    \caption{\label{fig:coordinates}Notation for the location and orientation of each bacterium on the surface of the particle.  The   $i$-th bacterium is located at  $\mathbf{x}_i=R\mathbf{e}_{r_i}$ where $R$ is the radius of the
    particle and $\mathbf{e}_{r_i}$ is the radial unit vector defined by the polar and azimuth angles $\vartheta_i$ and $\varphi_i$ with respect to the body frame $\{x,y,z\}$. The orientation of the bacterium relative to the bead
    surface is given by the unit vector $\mathbf{p}_{i}$, which is the radial unit vector defined by the polar and azimuth angles $\alpha_i$ and $\beta_i$ with respect to the local spherical coordinate system
    $\{\mathbf{e}_{\vartheta_i},\mathbf{e}_{\varphi_i}, \mathbf{e}_{r_i}\}$. Each bacterium is assumed to push on the fluid along $\mathbf{p}_{i}$ and thus to exert a force on the particle along $-\mathbf{p}_{i}$.}
   \end{figure}
   Each cell is assumed to perform its own, independent run-and-tumble dynamics. The reaction torques in the running and tumbling states have magnitudes $|\mathbf{M}_{_R}|\simeq 0.7\, \text{pN}\mu \text{m}$ and
   $|\mathbf{M}_{_T}|\simeq 0.4\, \text{pN}\mu \text{m}$ respectively \cite{zhuang2017propulsion}. These torques can be neglected if we focus on  sufficiently large   particles such that 
   $|\mathbf{M}_{_{R,T}}|\sqrt{\mean{\sin^2{\alpha}}_\alpha}\ll fR$, where $f\sim 0.3-0.48\,\text{pN}$ is the average propulsive force exerted by each bacterium \cite{Behkam2008, zhuang2017propulsion} and  the angle $\alpha$ defined
   in Fig.~\ref{fig:coordinates} denotes the orientation of the bacterium flagella bundle with respect to the radial direction. In other words, for large particles, the reorientation of the swimmer is dominated by the moment-arm torque
   induced by the  propulsive forces of the cells. Finally, thermal noise can also be neglected as it induces typical forces three orders of magnitude smaller than the propulsive forces from the bacteria.
   We model therefore each bacterium   as a two-state machine which exerts a force of magnitude $f$ when running and no force when tumbling. The transition between the running and the tumbling states is modelled as a continuous time
   Markov chain with transition rates $\lambda_{_R}$ and $\lambda_{_T}$   from the tumbling to the running state and vice-versa, respectively (see Fig.~\ref{fig:r-n-t}). The values for \textit{E.~coli} in homogeneous environments
   (no chemical gradients) are  \mbox{$\lambda_{_R}\simeq10$ s$^{-1}$} and \mbox{$\lambda_{_T}\simeq10/9$ s$^{-1}$}. Furthermore, the swimming motion occurs at low Reynolds number $Re\simeq10^{-4}$ \cite{BLOCK1982215}. The linear and angular
   velocities of the  micro-swimmer, denoted by $\mathbf{V}$ and $\boldsymbol{\omega}$, are obtained by force and moment balance using Stokes law
   \begin{align}
    \label{eq:Force}
    \mathbf{F}&=6\pi\mu R\mathbf{V}=-f\sum_{i=1}^{N}{\epsilon_{i}\mathbf{p}_{i}},\\
    \label{eq:Torque}
    \mathbf{M}&=8\pi\mu R^3\boldsymbol{\omega}=-fR\sum_{i=1}^{N}{\epsilon_{i}\mathbf{e}_{r_i}\times\mathbf{p}_{i}},
   \end{align}
   \noindent where $\mu$ is the dynamic viscosity of the fluid,  $\mathbf{F}$ and $\mathbf{M}$ are the total force and torque applied  by the surface-attached cells, while $\mathbf{e}_{r_i}$ and $\mathbf{p}_i$ are unit vectors which
   determine the position and orientation of the $i$-th bacterium (see  Figs.~\ref{fig:micro_swimmer}b and \ref{fig:coordinates}).  Specifically, each bacterium is assumed to push on the fluid along the direction $\mathbf{p}_{i}$ of
   its bundle of flagellar filaments and thus to exert a force on the particle along $-\mathbf{p}_{i}$. In Eqs.~\eqref{eq:Force} and \eqref{eq:Torque}, the stochastic variable $\epsilon_i$ determines the state of the $i$-th bacterium
   with $\epsilon_i=1$ when running and $\epsilon_i=0$ when tumbling. The position and orientation of the micro-swimmer evolve in time according to
   \begin{align}
    \label{eq:Position}
    \dot{\mathbf{X}}=\mathbf{V},\\
    \label{eq:Orientation}
    \dot{\mathbf{n}}=  \boldsymbol{\omega}\times\mathbf{n},
   \end{align}
   \noindent where the  dot denotes a derivative with respect to time and $\mathbf{n}$ is a body-fixed unit vector. We integrate Eq.~\eqref{eq:Position} numerically using an Euler method and Eq.~\eqref{eq:Orientation} using the mid point
   method presented in Refs.~\cite{Simo1995, ZUPAN2011723} with a time step $\text{d}t=0.01$ s. We consider $10^4$ different configurations, sampling the geometrical angles  $\vartheta_i$, $\varphi_i$, $\alpha_i$ and $\beta_i$, which describe
   the location and orientation of the cells on the bead surface (see Fig.~\ref{fig:coordinates}), from uniform distributions in the intervals $[0,\pi]$, $[0,2\pi)$, $[\alpha_{\text{max}},\alpha_{\text{min}}]$ and $[0,2\pi)$ respectively.
  The maximum and minimum deviation angles from the radial direction are  $\alpha_{\text{min}}=30^\circ$ and $\alpha_{\text{max}}=85^\circ$ following  Ref.~\cite{arabagi2011modeling}.
  \subsection{\label{subsec:r-n-t_prob} Run-and-tumble dynamics}
   If we denote by $p_{_R}(t)$ and $p_{_T}(t)$ the probabilities of finding a particular bacterium in the running and the tumbling states respectively, the master equation for the run-and-tumble dynamics is given by
   \begin{equation}
    \label{eq:prob_1}
    \left[\begin{array}{c}\dot{p}_{_R} \\ \dot{p}_{_T}\end{array}\right]=\left[\begin{array}{cc} -\lambda_{_T}	& \lambda_{_R}\\ \lambda_{_T}	& -\lambda_{_R} \end{array}\right]\left[\begin{array}{c} p_{_R} \\ p_{_T}\end{array}\right],
   \end{equation}
   where $\lambda_{_R}$ and $\lambda_{_T}$ are the transitions rates defined earlier. 
   The system in Eq.~\eqref{eq:prob_1} can be easily integrated by considering the normalisation condition $p_{_R}+p_{_T}=1$, with the result 
   \begin{equation}
    \label{eq:prob_2}
    \left[ \displaystyle\begin{array}{c} p_{_R} \\ p_{_T}\end{array}\right]=\left[\begin{array}{c} \displaystyle \frac{\lambda_{_R}}{\lambda_{_R}+\lambda_{_T}}+Ae^{-\left(\lambda_{_R}+\lambda_{_T}\right)t} \\
    \displaystyle \frac{\lambda_{_T}}{\lambda_{_T}+\lambda_{_R}}-Ae^{-\left(\lambda_{_R}+\lambda_{_T}\right)t}\end{array}\right],
   \end{equation}
   \noindent where $A$ is a constant determined by the initial state of the bacterium. In particular, if the bacterium is running at time $t=0$ we have $p_{_R}(0)=1$, hence
   \begin{equation}
    \label{eq:prob_2_1}
    \left[\begin{array}{c} p_{_{RR}} \\ p_{_{TR}}\end{array}\right]=\left[\begin{array}{c} \displaystyle\frac{\lambda_{_R}+\lambda_{_T}e^{-\left(\lambda_{_R}+\lambda_{_T}\right)t}}{\lambda_{_R}+\lambda_{_T}} \\
    \displaystyle  \frac{\lambda_{_T}-\lambda_{_T}e^{-\left(\lambda_{_R}+\lambda_{_T}\right)t}}{\lambda_{_R}+\lambda_{_T}}\end{array}\right],
   \end{equation}
   \noindent where we denote by $p_{_{RR}}(t)$, $p_{_{TR}}(t)$ the probability for a bacterium to be running or tumbling at time $t$ given that it was running at time $t=0$. 
   Furthermore, in the steady state, the results of Eq.~\eqref{eq:prob_2} reduce to
   \begin{equation}
    \label{eq:prob_3}
    \left[\begin{array}{c} p_{_R}^s \\ p_{_T}^s\end{array}\right]=\left[\begin{array}{c} \displaystyle\frac{\lambda_{_R}}{\lambda_{_R}+\lambda_{_T}} \\
    \displaystyle \frac{\lambda_{_T}}{\lambda_{_T}+\lambda_{_R}}\end{array}\right].
   \end{equation}
   In our numerical simulations, we assume that we have waited long enough so that the system has reached   steady state. We therefore determine the state of the variables $\epsilon_i$ by sampling a pseudo random number $r_i$
   from a uniform distribution and comparing it with $p_{_R}^s$ for the first step and $p_{_{RR}}(\text{d}t)$, $p_{_{RT}}(\text{d}t)$ for the following steps. Here $p_{_{RT}}$ is obtained from Eq.~\eqref{eq:prob_2} by taking $p_{_R}(0)=0$. In the
   steady state, the mean of the variables $\epsilon_i(t)$ is
   \begin{equation}
    \label{eq:prob_4}
    \mean{\epsilon_i(t)}=P(\epsilon_i(t)=1)=p_{_R}^s=\frac{\lambda_{_R}}{\lambda_{_R}+\lambda_{_T}},
   \end{equation}
   while the time autocorrelation is given by
   \begin{align}
    \label{eq:prob_5}
    \mean{\epsilon_i(t)\epsilon_j(s)}&=P(\epsilon_i(t)=1,\epsilon_j(s)=1)\nonumber\\
    &=P(\epsilon_j(s)=1|\epsilon_i(t)=1)p_{_R}^s.
   \end{align}
   If the bacteria behave independently from each other, the probability of the $j$-th bacterium being running at time $s$, given that the $i$-th bacterium is also running at time $t$, is $p_{_{RR}}(|s-t|)$ for $i=j$ and $p_{_R}^s$
   otherwise. Hence
   \begin{equation}
    \label{eq:prob_6} 
    \mean{\epsilon_i(t)\epsilon_j(s)}=\left(p_{_R}^s\right)^2\left(1+\delta_{ij}\frac{\lambda_{_T}}{\lambda_{_R}}e^{-\left(\lambda_{_R}+\lambda_{_T}\right)|s-t|}\right).
   \end{equation}
   We will use Eqs.~\eqref{eq:prob_4} and \eqref{eq:prob_6} to evaluate the MSD bellow.
  \subsection{\label{subsec:single_chemo} Chemotaxis of a single bacterium}
   Bacteria such as \textit{E.~coli} are able to navigate through chemical gradients by using  an inhibition of the mechanism that inverts the polarity of their bacterial motor when the cell swims in a favourable direction
   \cite{Berg2004}. Polarity inversion is responsible for  tumbling events, and therefore runs are extended when swimming up (resp.~down) the chemoattractant (resp.~repellent) gradient \cite{SOURJIK2012262}. To investigate whether the
   individual chemotaxis of single cells contributes collectively to chemotaxis of the micro-swimmer, we use the simplest mathematical model that describes chemotaxis of a single bacterium. Specifically, we model each organism as
   equipped with an internal sensor that responds to temporal variations in the chemical concentration of its environment by modifying the tumbling rate, $\lambda_{_{T}}$, i.e.~the expected rate at which running cells stop running and
   transition to a tumble~\cite{Schnitzer1993}. For weak chemical concentrations $c(t)$ sensed by the bacterium, the variation in $\lambda_{_{T}}$ is captured by a linear response as \cite{deGennes2004}
   \begin{equation}
    \label{eq:chem_1}
    \lambda_{_{T}}(t)=\lambda_{0}\left(1-\int_{-\infty}^{t}{K(t-t')c(t')\text{d}t'}\right),
   \end{equation}
   \noindent where $\lambda_0$ is the mean tumbling rate when $c=c_0$ is constant. The function $K(t)$ is a memory kernel which describes the response and has been measured for \textit{E.~coli} by Segall et
   al.~\cite{Segall1986}. For a constant concentration gradient, the variation in the concentration experienced by the bacterium will depend on its position as follows
   \begin{equation}
    \label{eq:chem_2}
    c(t)=c_0+\mathbf{r}(t)\cdot\nabla c=c_0+\nabla c\cdot\int_{0}^{t}{\mathbf{v}(t')\text{d}t'},
   \end{equation}
   \noindent where $c_0$ is a constant background concentration and $\mathbf{r}(t)$ and $\mathbf{v}(t)$ are, respectively, the position and velocity of the bacterium at time $t$  measured from the beginning of the last running event.
   The memory kernel $K(t)$ satisfies the condition of adaptability \cite{Schnitzer1993}
   \begin{equation}
    \label{eq:chem_4}
    \int_{0}^{\infty}{K(t)\text{d}t}=0,
   \end{equation}
   \noindent which ensures $\lambda_{_T}(t)=\lambda_0$ when $c$ is kept constant. Inserting Eqs.~\eqref{eq:chem_2} into Eq.~\eqref{eq:chem_1} we obtain the tumbling rate as
   \begin{equation}
    \label{eq:chem_5}
    \lambda_{_T}(t)=\lambda_{0}\left[1-\nabla c\cdot\int_{0}^{\infty}{\int_{0}^{t-s}{K(s)\mathbf{v}(s')\text{d}s'}\text{d}s}\right].
   \end{equation}
   For computational convenience, we consider in our numerical simulations the simplest possible adaptive kernel. Specifically, we take only two impulses localised at $t_1$ and $t_2$ with
   $t_1<t_2$, and therefore write $K(t)=\kappa\left[\delta(t-t_1)-\delta(t-t_2)\right]$, where $\kappa$ is the magnitude of the response. This models a cell which compares concentrations at two different, specific times in the past,
   and changes its tumbling rate accordingly. Substitution of this kernel into Eq.~\eqref{eq:chem_5} yields
   \begin{equation}
    \label{eq:chem_6}
    \lambda_{_T}(t)=\lambda_{0}\bigg \{1-\kappa\nabla c\cdot \big[\mathbf{r}(t-t_1)-\mathbf{r}(t-t_2) \big]\bigg\}.
   \end{equation}
   Note that, since any function can be expressed as a superposition of impulses,     it is possible to treat more general chemotactic responses (such as those proposed by Clark et al.~\cite{Clark2005} or Celani et
   al.~\cite{Celani1391}) in a similar manner by approximating $K(t)$ by a sum of delta functions, \mbox{$K(t)\simeq\sum\kappa_j\delta(t-t_j)$}, where $t_j$'s are delay times and where the $\kappa_j$'s determine the intensity of the response
    subject to \mbox{$\sum{\kappa_j}=0$}.
  \subsection{\label{subsect:chem_micro_num} Chemotaxis of bacteria-driven micro-swimmers}
   Chemotaxis is included numerically in our model by modifying the tumbling rate of each bacterium according to Eq.~\eqref{eq:chem_6}. We make the simplifying assumption that the particle is permeable to the chemical so that the chemical gradient is not perturbed by its presence.
     In general, we should evaluate $\mathbf{r}(t)$ in Eq.~\eqref{eq:chem_6} as the position of each cell. However,
   if we consider shallow gradients such that $\kappa|\nabla c|R\ll1$, then it is appropriate to neglect variations in concentration along the surface of the particle and substitute the position of the centre of the micro-swimmer for
   $\mathbf{r}(t)$ in Eq.~\eqref{eq:chem_6}.
   
   We run the simulation from the previous sections for an ensemble of $10^4$ beads, now evaluating the tumbling rate at each step using Eq.~\eqref{eq:chem_6} with $\mathbf{r}(t)$ taken to be the position of the centre of the
   bead. We further take the position of the two impulses in the kernel to coincide with the maximum and minimum of the response kernel measured by  Segall et al.~\cite{Segall1986}, \textit{i.e.}~\mbox{$t_1\simeq 1\, \text{s}$} and
   \mbox{$t_2\simeq 3\, \text{s}$}. The intensity of the response is taken as \mbox{$\kappa=|K|_{max}\times 1\,\text{s} \simeq0.3\,\mu \text{M}^{-1}$} \cite{Celani1391}.  As we require the second term in brackets in Eq.~\eqref{eq:chem_6} to remain small
   and $\mathbf{r}(t)\sim U_e (t_2-t_1)$, where $U_e\sim 10\, \mu\text{m}/\text{s}$ is the swimming speed of the micro-swimmers \cite{Behkam2008}, we choose \mbox{$\kappa|\nabla c|\sim 10^{-3}-10^{-2}\,\mu \text{m}^{-1}$}, which corresponds to concentration gradients
   $|\nabla c|\sim 3\times10^{-3}-3\times10^{-2}\,\text{mM}/\text{mm}$, so that $\kappa|\nabla c|R\sim10^{-2}-10^{-1}$ and the tumbling rate is reduced by approximately one tenth. The chemical gradient is set up along the $z$ direction.
 \section{\label{sec:results} Theoretical and numerical results}
  We start this section with the description of the three-dimensional trajectories of the micro-swimmers as obtained numerically and we observe that for long time-scales these become three-dimensional random walks. We then introduce a coarse-grained model to
  describe the diffusive behaviour of the micro-swimmers. We define the rotational diffusion coefficient, $D_r$, the effective speed, $U_e$, the effective diffusion coefficient, $D_e$ and   derive analytical expressions for each one of
  them. Next, we show that in the   presence of a non uniform chemical concentration field, to which the bacteria respond chemotactically, the micro-swimmers perform a biased random walk and we quantify their response in terms of a
  drift speed $v_d$, for which we derive an analytical expression. Throughout this section, we validate the analytical results against our numerical simulations.
  \subsection{\label{subsec:trajectories} Three-dimensional trajectories of bacteria-driven micro-swimmers}
   \begin{figure}
    \centering
    \includegraphics[clip, trim=0cm 2.75cm 0cm 3.5cm, width=0.9\columnwidth]{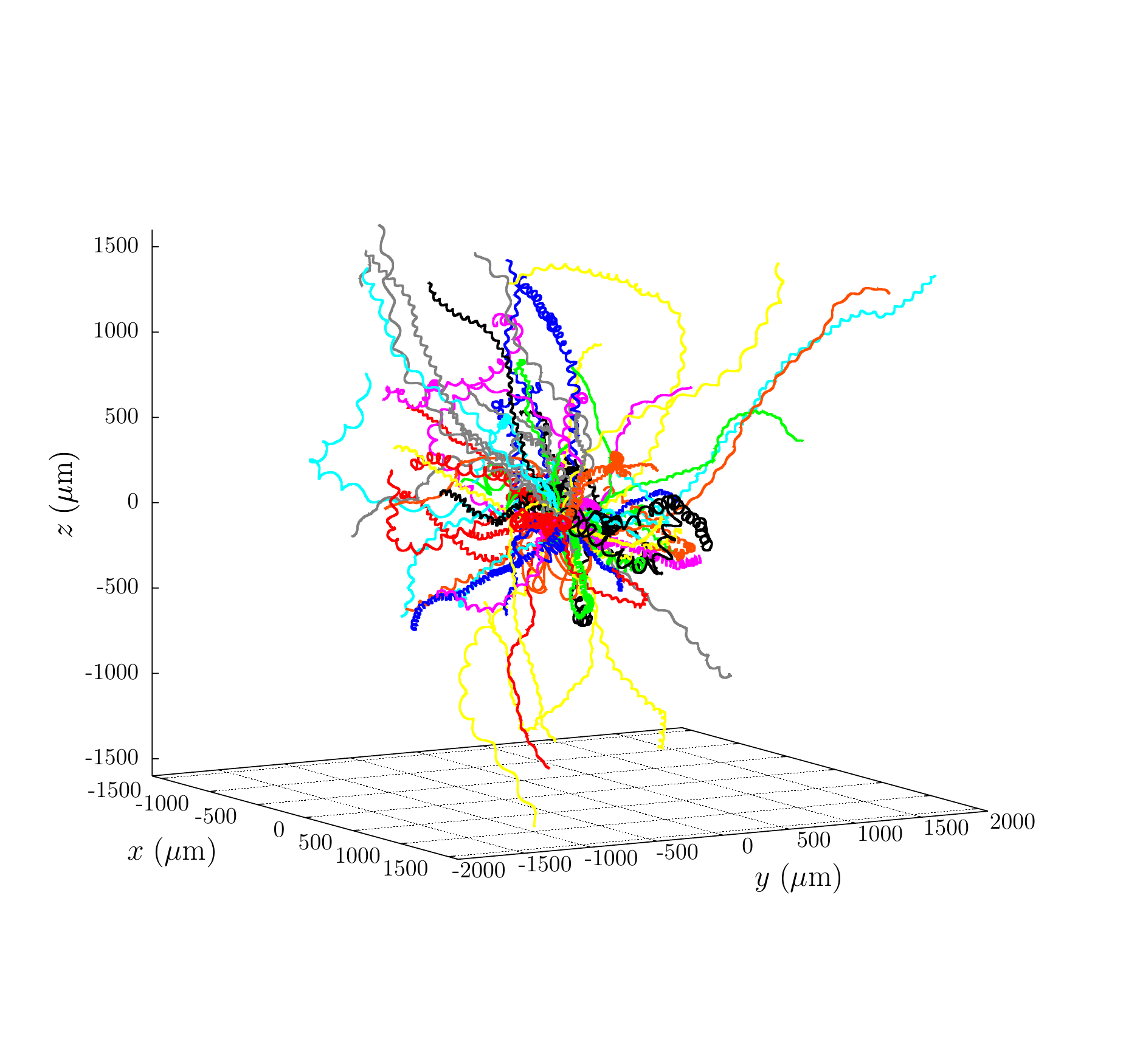}
    \caption{\label{fig:helices_2} Typical trajectories of bacteria-driven micro-swimmers  of radius $R=10\,\mu$m with bacteria density $\rho=1/12\; \,\mu \text{m}^{-2}$, individual propulsive forces $f=0.48$ pN, and transition rates $\lambda_{_T}^{-1}=0.9$ s and
    $\lambda_{_R}^{-1}=0.1$ s. The simulation time is $t=5/(3D_r)\simeq 224\,\text{s}$, where $D_r$ is given by Eq.~\eqref{eq:rot_diff_11}. One hundred different realisations are shown.}
   \end{figure}
   We present first the results in isotropic environments, i.e.~constant concentration field and   thus constant tumbling rate. Typical trajectories for the centre of the particle, resulting from numerical integration
   of equations \eqref{eq:Position} and \eqref{eq:Orientation}, are shown in Fig.~\ref{fig:helices_2}. At short time scales, the applied force and torque are nearly constant and the resulting trajectories are noisy helices.
   For a given distribution of attached bacteria, the average force and torque acting on the particles in the steady state are given by
   \begin{align}
    \label{eq:Force_1}
    \mean{\mathbf{F}}_\epsilon&=-f\sum_{i=1}^{N}{\mean{\epsilon_i}_\epsilon\mathbf{p}_i}=-fp_{_R}^s\sum_{i=1}^{N}{\mathbf{p}_i},\\
    \label{eq:Torque_1}
    \mean{\mathbf{M}}_\epsilon&=-fR\sum_{i=1}^{N}{\mean{\epsilon_i}_\epsilon\left(\mathbf{e}_{r_i}\times\mathbf{p}_i\right)}=-fRp_{_R}^s\sum_{i=1}^{N}{\left(\mathbf{e}_{r_i}\times\mathbf{p}_i\right)},
   \end{align}
   \noindent where $\mean{\cdot}_\epsilon$ denotes the average over the probability distribution Eq.~\eqref{eq:prob_2}. Defining the vectors $\mathbf{a}\equiv\sum{\mathbf{p}_i}$ and
   $\mathbf{b}\equiv\sum{\left(\mathbf{e}_{r_i}\times\mathbf{p}_i\right)}$, we can express the average radius and pitch of the trajectories as follows 
   \begin{align}
    \label{eq:rad_pitch_2a}
    \mathcal{R}&=\frac{4}{3}R^2\frac{\left|\mean{\mathbf{F}}_\epsilon\times\mean{\mathbf{M}}_\epsilon\right|}{\left|\mean{\mathbf{M}}_\epsilon\right|^2}=\frac{4}{3}R\frac{\left|\mathbf{a}\times\mathbf{b}\right|}{\left|\mathbf{b}\right|^2},\\
    \label{eq:rad_pitch_2b}
    \mathcal{P}&=\frac{8\pi}{3}R^2\frac{\left|\mean{\mathbf{F}}_\epsilon\cdot\mean{\mathbf{M}}_\epsilon\right|}{\left|\mean{\mathbf{M}}_\epsilon\right|^2}=
    \frac{8\pi}{3}R\frac{\left|\mathbf{a}\cdot\mathbf{b}\right|}{\left|\mathbf{b}\right|^2}\cdot
   \end{align}
   In contrast, at longer times the trajectories become three-dimensional random walks and their diffusive behaviour determines the motility properties of the micro-swimmers. It is well known in the theory of Brownian motion that for a
   random walk governed by rotational diffusion, the MSD is given by \cite{Chandrasekhar1943}
   \begin{equation}
    \label{eq:MSD_1}
    \mean{r^2(t)}=2\tau_r^2U_e^2\left(\frac{t}{\tau_r}+e^{-t/\tau_r}-1\right),
   \end{equation}
   \noindent where the angle brackets $\mean{\cdot}$ denote ensemble average, $U_e$ is the effective speed at which the bead moves and $\tau_r$ is the orientation correlation time, which is the time scale of decay for the orientation
   correlation function, $\mean{\mathbf{n}(0)\cdot\mathbf{n}(t)}$. In other words, $\tau_r$ is the time it takes the micro-swimmer to forget its initial orientation. For short time scales, that is $t\ll\tau_r$, Eq.~\eqref{eq:MSD_1}
   reduces to $\mean{r^2(t)}=U_e^2t^2$, which represents ballistic motion. On the other hand, for time scales such that $\tau_r\ll t$ the MSD is linear in $t$, a dependence typical in diffusion processes. The constant of proportionality
   is   the effective diffusion coefficient which is defined in three dimensions by
   \begin{equation}
    \label{eq:diff_eff_0}
    6D_e\equiv\lim_{t\rightarrow\infty}{\frac{\mean{r^2(t)}}{t}}=2U_e^2\tau_r.
   \end{equation}
   Since $D_e$ is a macroscopic property of the micro-swimmers, it is likely that a simplified description of the trajectories ignoring the microscopic details would still lead to the  same result. Furthermore, ignoring the fine
   structure of the driving mechanism will render the chemotaxis analysis more tractable (see Sec.~\ref{sec:chemotaxis}). Inspired by this we propose a coarse grained model as described in the next section. We will show that the
   micro-swimmers MSD is described by Eq.~\eqref{eq:MSD_1} and we will derive analytical expressions for $U_e$, $\tau_r$ and $D_e$.
   \subsection{\label{subsec:coarse} Coarse-grained modelling}
   \begin{figure}
    \centering
    \includegraphics[clip, trim=0.2cm 3.4cm 0.3cm 3.8cm, width=0.95\columnwidth]{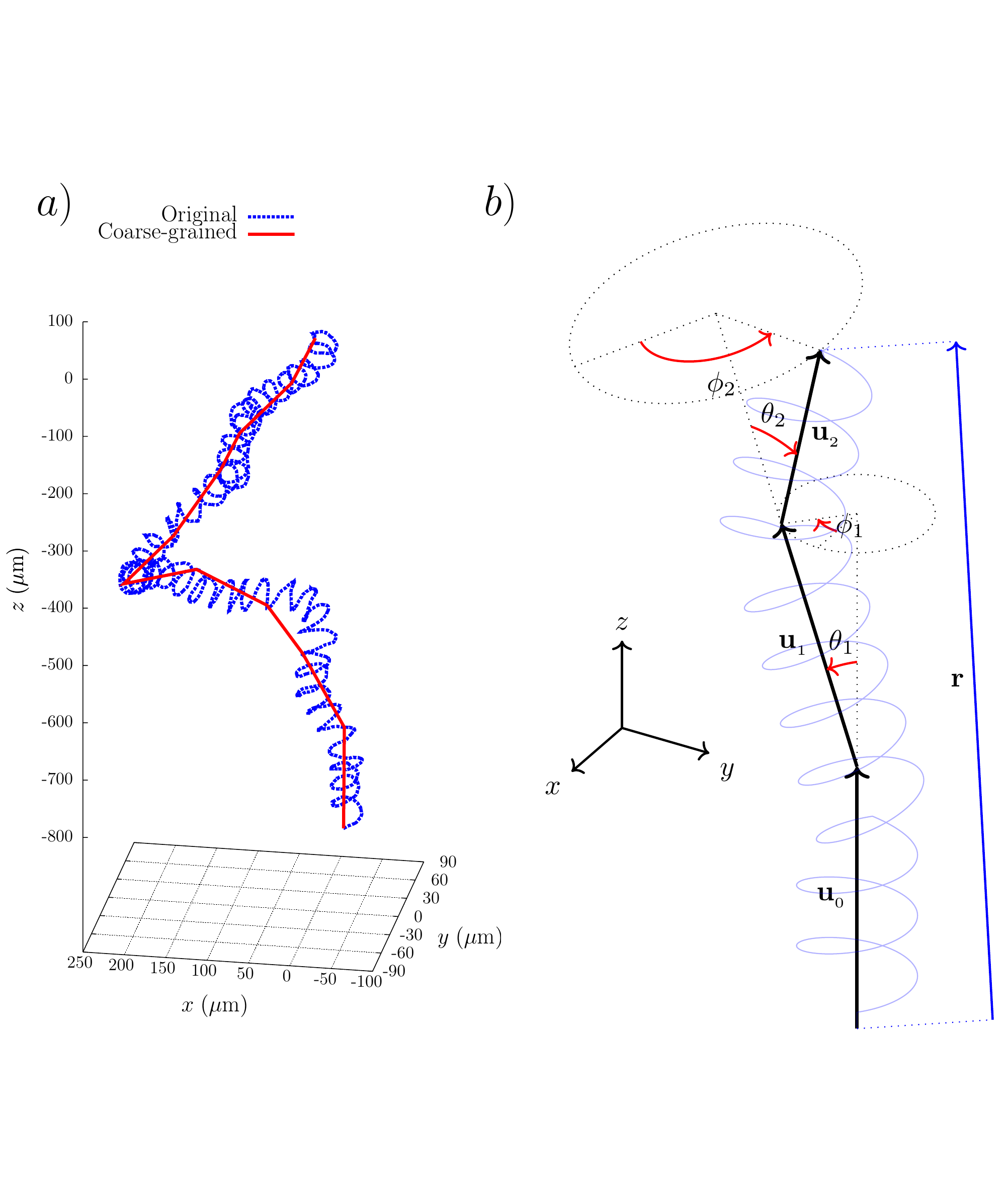}
    \caption{\label{fig:helices_1} 
    Coarse graining the trajectories of particles. (\textit{a}) Typical trajectory of a bacteria-driven micro-swimmer with parameters: radius $R=10\,\mu\text{m}$, cell density $\rho=1/12\; \,\mu \text{m}^{-2}$, propulsive force $f=0.48$ pN, and transition rates $\lambda_{_T}^{-1}=0.9$ s and $\lambda_{_R}^{-1}=0.1$ s.
    The simulation time is \mbox{$t=5/(3D_r)\simeq 224\,\text{s}$}, where $D_r$ is given by Eq.~\eqref{eq:rot_diff_11}. The path of the swimmer is plotted in dotted blue line and can be approximated by a coarse-grained trajectory in the form of a random walk (solid red line). 
    (\textit{b}) The coarse-grained trajectory is analogous to a freely rotating chain with equal-length links. The angle between two 
    consecutive paths $\hat{\mathbf{u}}_{i-1}$ and $\hat{\mathbf{u}}_i$ is denoted by $\theta_i$ and the internal angle of rotation, denoted by $\phi_i$, is the angle between the planes generated by $\{\hat{\mathbf{u}}_{i-2},
    \hat{\mathbf{u}}_{i-1}\}$ and $\{\hat{\mathbf{u}}_{i-1}, \hat{\mathbf{u}}_{i}\}$.}
   \end{figure}
   We start by observing that the trajectories of the micro-swimmers such as those illustrated in Fig.~\ref{fig:helices_2} and Fig.~\ref{fig:helices_1}a consist of almost undisturbed helical paths interrupted by sudden changes
   in direction. We thus construct a coarse-grained model replacing every helical trajectory by a straight path along an average axis denoted by $\hat{\mathbf{u}}_k$ for the $k$th path  (see Fig.~\ref{fig:helices_1}b). Each straight
   path is travelled at a constant speed $U_e$ for an average time $T$. The resulting trajectory is therefore a chain of equal-length links. The polar and azimuthal angles between consecutive paths $\hat{\mathbf{u}}_{i-1}$ and
   $\hat{\mathbf{u}}_i$ are denoted by $\theta_i$ and $\phi_i$ respectively.   We further take  $\phi_i$ as uniformly distributed, due to homogeneity of the space, but we will assume that there is some persistence in the direction
   of motion, as can be observed in Fig.~\ref{fig:helices_2}.

   In polymer physics, a mathematically-identical setup is used in the  freely-rotating chain model of  a polymer and it is thus a textbook result that the velocity correlation function satisfies \cite{LOVELY1975477, Rubinstein2003, Yamakawa1997}
   \begin{align}
    \label{eq:poly_0}
    \mathcal{C}(t)&\equiv\mean{\mathbf{u}(0)\cdot\mathbf{u}(t)}=U_e^2\sum_{q=0}^{\infty}{P_q\mean{\hat{\mathbf{u}}_0\cdot\hat{\mathbf{u}}_q}_{\theta, \phi}}\nonumber\\
    &=U_e^2\sum_{q=0}^{\infty}{P_q\mean{\cos{\theta}}_{\theta}^{q}},
   \end{align}   
   \noindent where $\mean{\cos{\theta}}_\theta$ is the average cosine of the angle between consecutive paths. Since a change in direction on a trajectory is the result of several consecutive tumbling events of the bacteria, which are
   Poisson distributed, the probability $P_q$ of observing $q$ turning events in an interval of time $t$ must be Poisson distributed as well. Considering this and the fact that the turning rate is $T^{-1}$ we obtain
   \begin{align}
    \label{eq:poly_1}
    \mathcal{C}(t)&=U_e^2\sum_{q=0}^{\infty}{\frac{\left(t/T\right)^q}{q!}e^{-t/T}\mean{\cos{\theta}}_{\theta}^q}\nonumber\\
    &=U_e^2\exp{\left(-\frac{1-\mean{\cos{\theta}}_\theta}{T}t\right)}.
   \end{align}
   The MSD follows from a double integration in time of Eq.~\eqref{eq:poly_1}, leading to the  result 
   \begin{align}
    \label{eq:poly_2}
    \mean{r^2(t)}=\frac{2U_e^2T^2}{(1-\left\langle\cos{\theta}\right\rangle_\theta)^2}&\left(\frac{1-\mean{\cos{\theta}}_\theta}{T}t\right.\nonumber\\
    &\left.+\exp{\left[-\frac{(1-\left\langle\cos{\theta}\right\rangle_\theta)}{T}t\right]}-1\right).
   \end{align}
   Comparing with Eq.~\eqref{eq:MSD_1} we observe that the correlation time $\tau_r$ and the average flight time $T$ are related by
   \begin{align}
    \label{eq:poly_3}
    \tau_r=\frac{T}{1-\mean{\cos{\theta}}_\theta}\cdot
   \end{align}
   The average flight time $T$ can also be obtained from the definition of the Kuhn length as \cite{Rubinstein2003, Yamakawa1997}
   \begin{equation}
    \label{eq:poly_4}
    b\equiv\lim_{L\rightarrow\infty}{\frac{\mean{r^2(t)}}{L(t)}}=L(T),
   \end{equation}
   \noindent where $L(t)=\sqrt{3}U_e t$ is the maximum length of the trajectory (see Appendix \ref{appx:alpha_0}). Substitution of Eq.~\eqref{eq:poly_2} and Eq.~\eqref{eq:poly_3} into Eq.~\eqref{eq:poly_4} yields
   \begin{equation}
    \label{eq:poly_3_1}
    b=\frac{2U_e\tau_r}{\sqrt{3}}=\sqrt{3}U_e T,
   \end{equation}
   \noindent and therefore
   \begin{equation}
    \label{eq:poly_5}
    \tau_r=\frac{3}{2}T, \quad \mean{\cos{\theta}}_\theta=\frac{1}{3}\cdot
   \end{equation}
   From rotational diffusion \cite{Saragosti2012, Doi1996}, the correlation time is related to the rotational diffusion coefficient $D_r$, in $d$ spatial dimensions, by $\tau_r=[(d-1)D_r]^{-1}$. Therefore for our micro-swimmer
   $T^{-1}=3D_r$. The value of $D_r$ depends on the distribution of the torque that acts on the micro-swimmer, for example, reorientation due to thermal noise yields $D_r^T=k_BT/8\pi\mu R^3$ from the Stokes-Einstein relation
   \cite{Doi1996}. In the next section we will calculate the value of $D_r$ explicitly, considering the variance of the torque $\mathbf{M}$ given by Eq.~\eqref{eq:Torque}.
   \subsubsection{\label{subsubsec:rot_diff_coeff} Rotational diffusion coefficient}
    We now calculate the rotational diffusion coefficient for the particle by   considering the variance of its angular displacement.  The orientation of the micro-swimmer at time $t$ can be obtained from its initial orientation
    by applying the rotation matrix \mbox{$\mathbf{R}(t)=\exp{\left[\boldsymbol{\Gamma}(t)\right]}$}, where $\boldsymbol{\Gamma}$ is a matrix such that for any vector $\mathbf{v}$ we have $\boldsymbol{\Gamma}\mathbf{v}=\boldsymbol{\gamma}
    \times\mathbf{v}$, with $\boldsymbol{\gamma}$ the angle vector
    \begin{equation}
     \label{eq:rot_diff_5}
     \boldsymbol{\gamma}(t)=\int_0^t{\boldsymbol{\omega}(s)\text{d}s}.
    \end{equation}
    Given an orthonormal basis $\mathcal{B}=\{\mathbf{e}_i\}_{i=1}^3$, the components of $\boldsymbol{\gamma}$ on $\mathcal{B}$ satisfy $\mean{\cos{|\gamma_i|}}=e^{-D_r t}$, which in the Gaussian limit (thermal noise \cite{Saragosti2012})
    reduces to $\text{Var}{[\gamma_i]}=2D_r t$, where $\text{Var}[\cdot]$ denotes the variance. For large $N$ we may assume that $\boldsymbol{\gamma}$ is spherically uniform, thus we define the rotational diffusion coefficient as
    \begin{equation}
     \label{eq:rot_diff_4}
     6D_r\equiv\lim_{t\rightarrow\infty}{\frac{\left[\mean{|\boldsymbol{\gamma}|^2}-|\mean{\boldsymbol{\gamma}}|^2\right]}{t}},
    \end{equation}
    Substituting  Eq.~\eqref{eq:Torque} in Eq.~\eqref{eq:rot_diff_5} leads to
    \begin{equation}
     \label{eq:rot_diff_6}
     \boldsymbol{\gamma}=-\frac{f}{8\pi\mu R^2}\int_0^t{\sum_{i=1}^{N}{\epsilon_i(s)\left(\mathbf{e}_{r_i}(s)\times\mathbf{p}_i(s)\right)}\text{d}s}\equiv\int_0^t{\sum_{i=1}^N\epsilon_i(s)\mathbf{c}_i(s)},
    \end{equation}
    \noindent where Eq.~\eqref{eq:rot_diff_6} defines the vectors $\mathbf{c}_i(s)$. The variance of the angle vector for a fixed configuration of attached bacteria is then given by
    \begin{align}
     \label{eq:rot_diff_8}
     \text{Var}\left[\boldsymbol{\gamma}\right]&=\int_0^t{\int_0^t{\sum_{i,j}^N{\left[\mean{\epsilon_i(s_1)\epsilon_j(s_2)}_\epsilon-\left(p_{_R}^s\right)^2\right]\mathbf{c}_i(s_1)\cdot\mathbf{c}_j(s_2)}\text{d}s_2}\text{d}s_1}\nonumber\\
     &=\int_0^t{\int_{0}^t{\left[p_{_{R}}^sp_{_{RR}}(|s_2-s_1|)-\left(p_{_R}^s\right)^2\right]\sum_{i=1}^N{\mathbf{c}_i(s_1)\cdot\mathbf{c}_i(s_2)} \text{d}s_2}\text{d}s_1}\nonumber\\
     &=\int_0^t{\int_{0}^t{p_{_R}^sp_{_T}^se^{-(\lambda_{_R}+\lambda_{_T})|s_2-s_1|}\sum_{i=1}^N{\mathbf{c}_i(s_1)\cdot\mathbf{c}_i(s_2)} \text{d}s_2}\text{d}s_1}.
    \end{align}
    For large $N$ we may replace the sum by $N$ times the average over configurations
    \begin{align}
     \label{eq:rot_diff_9_0_0}
     \sum_{i=1}^N{\mathbf{c}_i(s_1)\cdot\mathbf{c}_i(s_2)}&\simeq N\left(\frac{f}{8\pi\mu R^2}\right)^2\mean{\left(\mathbf{e}_{r_i}\times\mathbf{p}_i\right)_{s_1}\cdot\left(\mathbf{e}_{r_i}\times\mathbf{p}_i\right)_{s_2}}_{\mathbf{c}_i},
    \end{align}
    \noindent where the subscript $\mathbf{c}_i$ denotes that the average is taken over the distributions of $\mathbf{e}_{r_i}$ and $\mathbf{p}_i$, which is equivalent to the average over the angles $\vartheta_i$, $\varphi_i$, $\alpha_i$
    and $\beta_i$. As the positions and orientations of the bacteria are identically and independently distributed, the correlation function on the right hand side of  Eq.~\eqref{eq:rot_diff_9_0_0} does not depend on $i$, hence
    \begin{align}
     \label{eq:rot_diff_9_0_1}
     \mean{\left(\mathbf{e}_{r}\times\mathbf{p}\right)_{s_1}\cdot\left(\mathbf{e}_{r}\times\mathbf{p}\right)_{s_2}}_{\mathbf{c}}&=N\sum_{i, j}{\mean{\left(\mathbf{e}_{r_i}\times\mathbf{p}_i\right)_{s_1}
     \cdot\left(\mathbf{e}_{r_j}\times\mathbf{p}_j\right)_{s_2}}_{\mathbf{c}}}\nonumber\\
     =N\mean{\mathbf{b}(s_1)\cdot\mathbf{b}(s_2)}_{\mathbf{c}}&=N\mean{|\mathbf{b}|^2}_{\mathbf{c}}\mean{\hat{\mathbf{b}}(s_1)\cdot\hat{\mathbf{b}}(s_2)}_{\mathbf{c}}.
    \end{align}
    Remembering that $\mathbf{b}$ defines the direction of motion in our coarse-grained model, we have
    \begin{equation}
     \label{eq:rot_diff_9_1}
     \mean{\hat{\mathbf{b}}(s_1)\cdot\hat{\mathbf{b}}(s_2)}_{\mathbf{c}}=\frac{1}{U_e^2}\mathcal{C}(|s_2-s_1|)=e^{-2D_r|s_2-s_1|},
    \end{equation}
    \noindent from Eq.~\eqref{eq:poly_1}. On the other hand, since $\alpha_i$ is the angle between $\mathbf{e}_{r_i}$ and $\mathbf{p}_i$ and the bacterial distributions are independent and identical, then
    $\mean{|\mathbf{b}|^2}_\mathbf{c}=N\mean{\sin^2{\alpha}}_{\alpha}$. Therefore
    \begin{equation}
     \label{eq:rot_diff_9}
     \sum_{i=1}^N{\mathbf{c}_i(s_1)\cdot\mathbf{c}_i(s_2)}\simeq N\left(\frac{f}{8\pi\mu R^2}\right)^2\mean{\sin^2{\alpha}}_\alpha e^{-2D_r |s_2-s_1|}.
    \end{equation}
    Note that, in doing this approximation, we are ignoring the periodic component of the correlation function of the vectors $\mathbf{c}_i$. As it can be seen in Fig.~\ref{fig:vel_corr} this   decays on a time scale shorter
    than $\tau_r$ and, since   we are interested on the diffusive behaviour of the micro-swimmer,  we can ignore it\footnote{The torque $\mathbf{M}$ fluctuates around its mean $\mean{\mathbf{M}}_\varepsilon\propto\mathbf{b}$ on a time-scale in
    the order of $\lambda_{_T}^{-1}$ which as seen below is much smaller than $\tau_r$.}. A substitution of Eq.~\eqref{eq:rot_diff_9} into Eq.~\eqref{eq:rot_diff_6} yields
    \begin{align}
     \label{eq:rot_diff_10}
     \text{Var}\left[\boldsymbol{\gamma}\right]&=\frac{2Nf^2\mean{\sin^2{\alpha}}_\alpha}{\left(8\pi\mu R^2\right)^2}\int_0^t{\int_{s_1}^t{p_{_R}^sp_{_T}^se^{-(\lambda_{_R}+\lambda_{_T}+2D_r)(s_2-s_1)} \text{d}s_2}\text{d}s_1}\nonumber\\
     &=\frac{2Nf^2\mean{\sin^2{\alpha}}_\alpha}{\left(8\pi\mu R^2\right)^2}p_{_R}^sp_{_T}^s\left(\frac{t}{\lambda_{_R}+\lambda_{_T}+2D_r}\right)\nonumber\\
     &+\frac{2Nf^2\mean{\sin^2{\alpha}}_\alpha}{\left(8\pi\mu R^2\right)^2} p_{_R}^sp_{_T}^s\left(\frac{e^{-(\lambda_{_R}+\lambda_{_T}+2D_r)t}-1}{(\lambda_{_R}+\lambda_{_T}+2D_r)^2}\right).
    \end{align}
    We expect the rotational diffusion time scale to be larger than the running and tumbling times, \textit{i.e.}~to be in the limit $D_r\ll (\lambda_{_R}+\lambda_{_T})$,
    which  will be  verified later. In this limit, Eq.~\eqref{eq:rot_diff_10} simplifies to
    \begin{equation}
     \label{eq:rot_diff_11}
     D_r=\left(\frac{f}{8\pi\mu R^2}\right)^2\frac{N\mean{\sin^2{\alpha}}_0\lambda_{_R}\lambda_{_T}}{3(\lambda_{_R}+\lambda_{_T})^3}
     +\mathcal{O}\left(\frac{2D_r}{(\lambda_{_R}+\lambda_{_T})^4}\right).
    \end{equation}
    We note that $D_r$ vanishes for $\alpha_i=0$, and indeed if all the bacteria are oriented in the radial direction they produce no moment-arm torque from their propulsive force and there is no change in the orientation of the particle.
    To treat  this case one must include the effect of the reaction torques, the calculations are however more complicated since the direction of $\mathbf{M}_T$ varies in time with respect to the direction of $\mathbf{M}_R$. On the other
    hand, the effects of thermal diffusion (which had been neglected) can be easily included by adding $6Dt$ to the MSD and by replacing $D_r$ by \mbox{$D_r+D_r^T$} where \mbox{$D=k_BT/6\pi\mu R$} and \mbox{$D_r^T=k_BT/8\pi\mu R^3$}
    are the thermal translational and rotational diffusion coefficients respectively.
 
    In the particular case of a micro-swimmer   immersed in water we have \mbox{$\mu\simeq10^{-3}\,\text{pNs}/\mu\text{m}^2$} and with the parameters \mbox{$f\simeq 5\times10^{-1}\,\text{pN}$}, \mbox{$\lambda_{_R}\simeq10\,\text{s}^{-1}$},
    \mbox{$\lambda_{_T}\simeq1\text{s}^{-1}$} and \mbox{$\rho\simeq10^{-1}\mu\text{m}^{-2}$}, we obtain
    \begin{equation}
     D_r\simeq (2\,\mu\text{m}^2\text{s}^{-1})\mean{\sin^2{\alpha}}_{\alpha}R^{-2}.
    \end{equation}
    Therefore Eq.~\eqref{eq:rot_diff_11} is valid as long as we have \mbox{$(0.2\,\mu\text{m}^2)\mean{\sin^2{\alpha}}_{\alpha}R^{-2}\ll1$}. For a uniform distribution of angle of attachment in the range
    \mbox{$[\alpha_{\text{min}},\alpha_{\text{max}}]$}, \mbox{$\mean{\sin^2{\alpha}}_{\alpha}\simeq 0.71$}, hence the condition \mbox{$D_r\ll(\lambda_{_R}+\lambda_{_T})$} is accomplished provided that the particle is sufficiently large,
    \mbox{$R\gtrsim3\,\mu\text{m}$}. We now proceed to calculate the micro-swimmers effective speed.
   \subsubsection{\label{subsubsec:eff_vel} Effective speed}
    We define the effective speed as the ensemble average projection of the velocity along the angular velocity, that is
    \begin{equation}
     \label{eq:eff_vel_0}
     U_e^2\equiv\mean{\frac{\left(\mean{\mathbf{V}}_\epsilon\cdot\mean{\boldsymbol{\omega}}_\epsilon\right)^2}{|\mean{\boldsymbol{\omega}}_\epsilon|^2}}_\mathbf{p}=
     \left(\frac{fp_{_R}^s}{6\pi\mu R}\right)^2\mean{\left(\mathbf{a}\cdot\hat{\mathbf{b}}\right)^2}_\mathbf{p}
    \end{equation}
    \noindent where $\mean{\cdot}_\mathbf{p}$ denotes the average over configurations. For a large number of bacteria $N$, we may assume that the vector $\mathbf{a}$ is uniformly distributed on the sphere and therefore
    \mbox{$\mean{\mathbf{a}\cdot\hat{\mathbf{b}}}_\mathbf{p}=\mean{|\mathbf{a}|^2}_\mathbf{p}/3$}. Substitution in Eq.~\eqref{eq:eff_vel_0} yields
    \begin{equation}
     \label{eq:eff_vel_1}
     U_e^2=\left(\frac{fp_{_R}^s}{6\pi\mu R}\right)^2\frac{1}{3}\sum_{i,j}^{N}{\mean{\mathbf{p}_i\cdot\mathbf{p}_j}}_\mathbf{p}=\frac{N}{3}\left(\frac{f}{6\pi\mu R}\frac{\lambda_{_R}}{\lambda_{_R}+\lambda_{_T}}\right)^{2}
    \end{equation}
    \noindent where we have assumed that the bacteria behave independently of each other, that is \mbox{$\mean{\mathbf{p}_i\cdot\mathbf{p}_j}=\delta_{i,j}$}. Notice that $p_{_R}^s$ is the fraction of time that the bacteria exert a
    force on the sphere, hence the effective speed is simply the average swimming speed of each bacterium \mbox{$v_0\equiv fp_{_R}^s/(6\pi\mu R)$} multiplied by the average number of bacteria pushing along the axis of the helical path
    $\sqrt{N/3}$. Notably, for a fixed density of attached bacteria, $U_e$ is independent of the size of the micro-swimmer as $N\sim R^2$, this is in agreement with the numerical results of Arabagi et al. \cite{arabagi2011modeling}. On the
    other hand, $U_e$ is proportional to the square root of the number of bacteria, which is in agreement with the experimental results of Behkam et al.~\cite{Behkam2008}.
    
   \subsubsection{\label{subsubsec:diffusion} Long-time diffusion of bacteria-driven micro-swimmers}
    \begin{figure}
     \centering
     \includegraphics[clip, trim=.8cm 0cm 1.3cm 0.1cm, width=0.9\columnwidth]{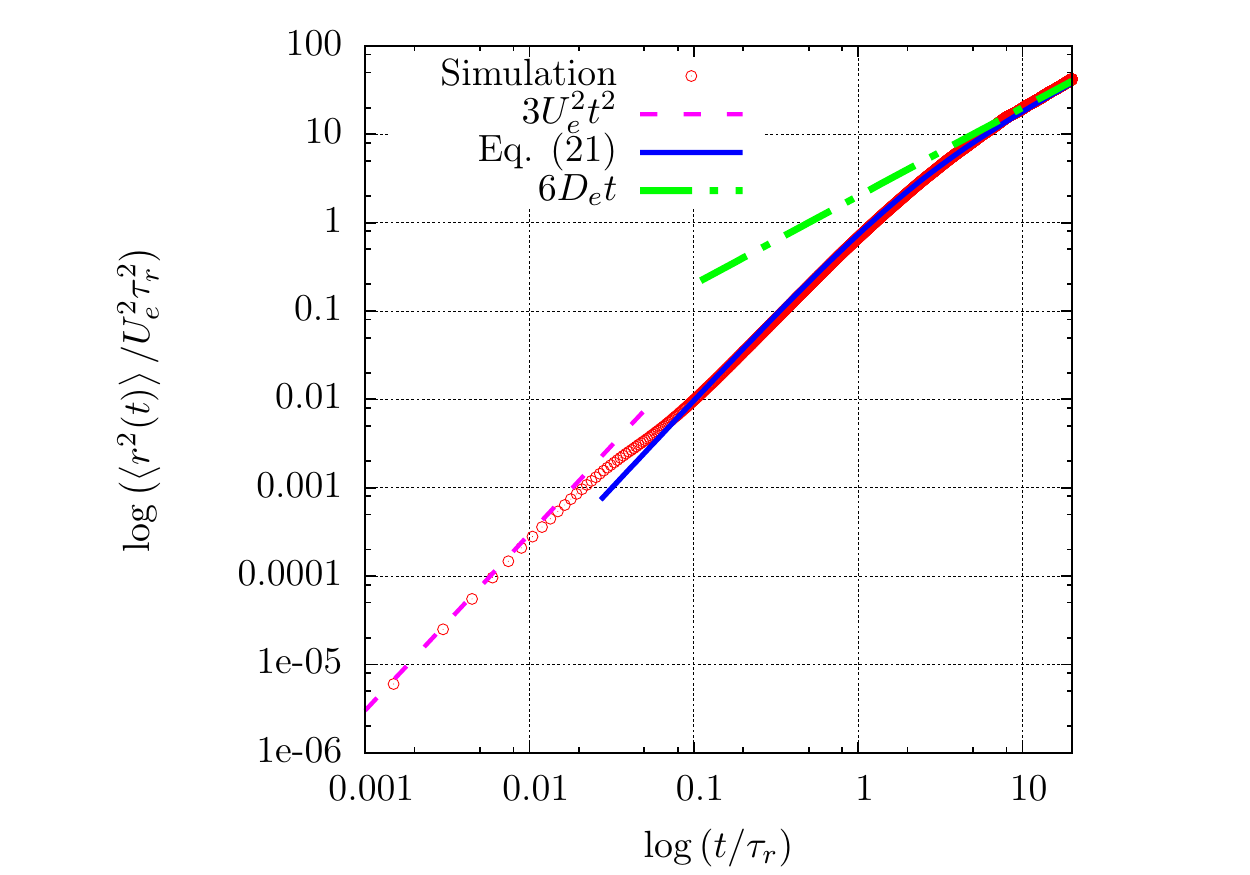}
     \caption{\label{fig:msd_log_plot}
     Mean squared displacement of bacteria-driven swimmers as a function of time, for fixed radius $R=10\,\mu\text{m}$ and density of cells $\rho=1/12\,\mu$m$^{-2}$. The parameters $f$,
     $\lambda_{_T}$ and $\lambda_{_R}$ are the same as in Fig.~\ref{fig:helices_2}. Circles:  mean values for $10^4$ different numerical realisations. Solid blue line; theory from  Eq.~\eqref{eq:MSD_1}. 
     Dashed dotted green line:      asymptotic diffusive behaviour $\mean{r^2}=6D_et$ and $D_e$ given by Eq.~\eqref{eq:diffusion_2}. 
     Dashed  magenta line:   instantaneous ballistic  evolution of the MSD,
     \mbox{$\mean{r^2}=3U_e^2t^2$}. Note that time is measured in units of $\tau_r\simeq68\,\text{s}$.}
    \end{figure}
    We now use the results above to obtain an analytical expression for the effective diffusion coefficient, $D_e$, defined by Eq.~\eqref{eq:diff_eff_0}. Substitution of Eqs.~\eqref{eq:rot_diff_11} and \eqref{eq:eff_vel_1} into
    Eq.~\eqref{eq:diff_eff_0} with $\tau_r^{-1}=2D_r$ leads to
    \begin{equation}
     \label{eq:diffusion_2}
     6D_e=\frac{16}{9}\frac{R^2\lambda_{_R}\left(\lambda_{_R}+\lambda_{_T}\right)}{\lambda_{_T}\left\langle\sin^2{\alpha}\right\rangle_\alpha}\cdot
    \end{equation}
    Our result makes two predictions with important consequences for experiments: (i) for a setup with a uniform density of surface-attached bacteria, we predict that the long-time diffusion coefficient increases with the square of the
    size of the particle; (ii) the diffusion constant is, perhaps surprisingly, independent of the value of the propulsive force exerted by each bacterium. Recall however that diffusion is obtained in the ``long-time" limit, which is
    defined as  $t \gg D_r^{-1}\sim f^{-2}$ and therefore a variation in the value of $f$ changes this limit. Note that when $\alpha_i=\pi/2$ for all $i$, we obtain a maximum value for the rotational diffusion and therefore a  minimum
    value of the  linear diffusion.  In contrast, when  $\alpha_i=0$ for all swimmers we obtain a singular expression for the diffusion coefficient, due to the fact that there is no reorientation as discussed in
    Sec.~\ref{subsubsec:rot_diff_coeff}. This limit is treated in Appendix \ref{appx:alpha_0}.
    \begin{figure}
     \centering
     \includegraphics[clip, trim=.8cm 0cm 1.3cm 0.1cm, width=0.9\columnwidth]{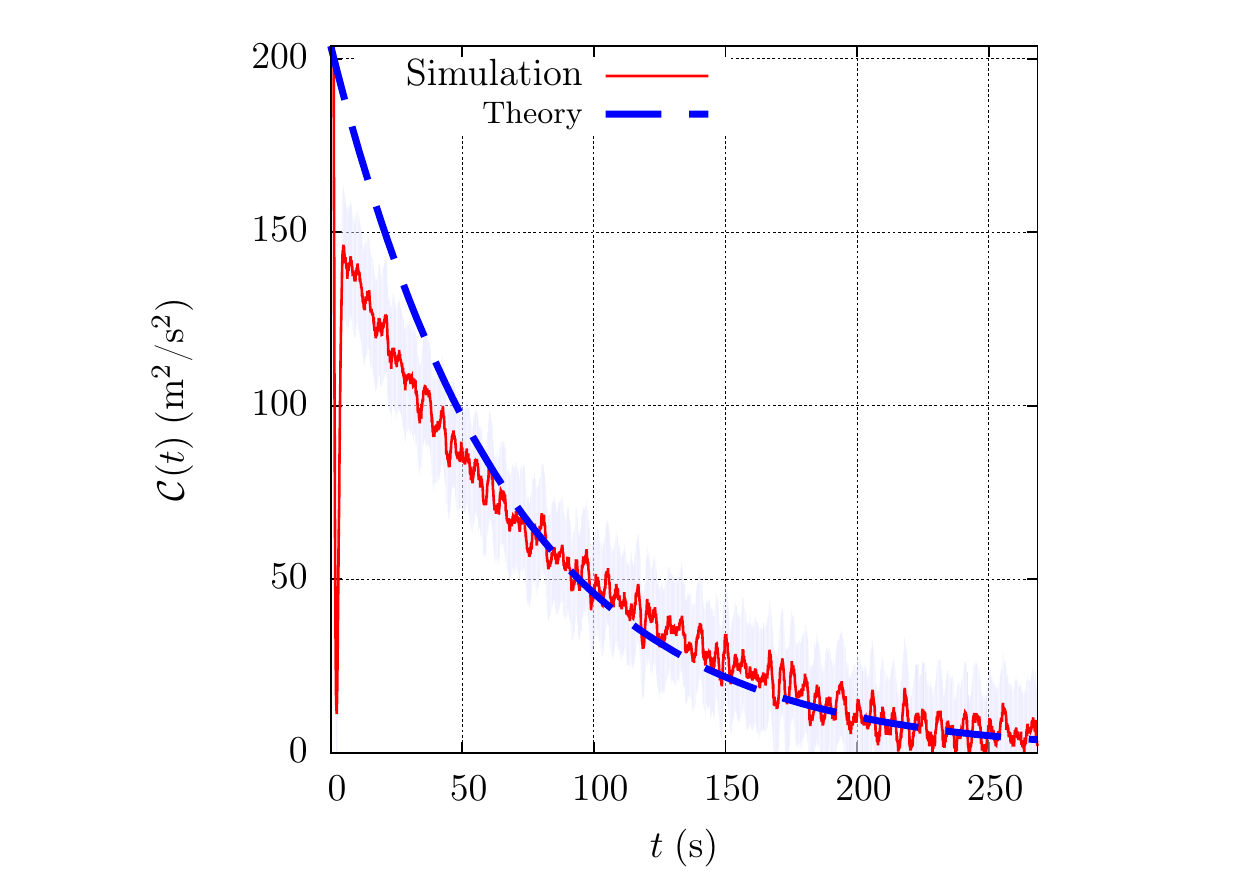}
     \caption{\label{fig:vel_corr} Velocity correlation function of a bacteria-driven swimmer of radius $R=10\,\mu\text{m}$ and density of bacteria $\rho=1/12\,\mu$m$^{-2}$. The parameters $f$, $\lambda_{_T}$ and $\lambda_{_R}$
     are the same as in Fig.~\ref{fig:helices_2}. The solid line shows the average over $10^4$ realisations and the shaded region the standard deviation. The dashed line is the theoretical prediction from Eq.~\eqref{eq:poly_1}.}
    \end{figure}
    \begin{figure}
     \centering
     \includegraphics[clip, trim=.8cm 0cm 1.3cm 0.1cm, width=0.9\columnwidth]{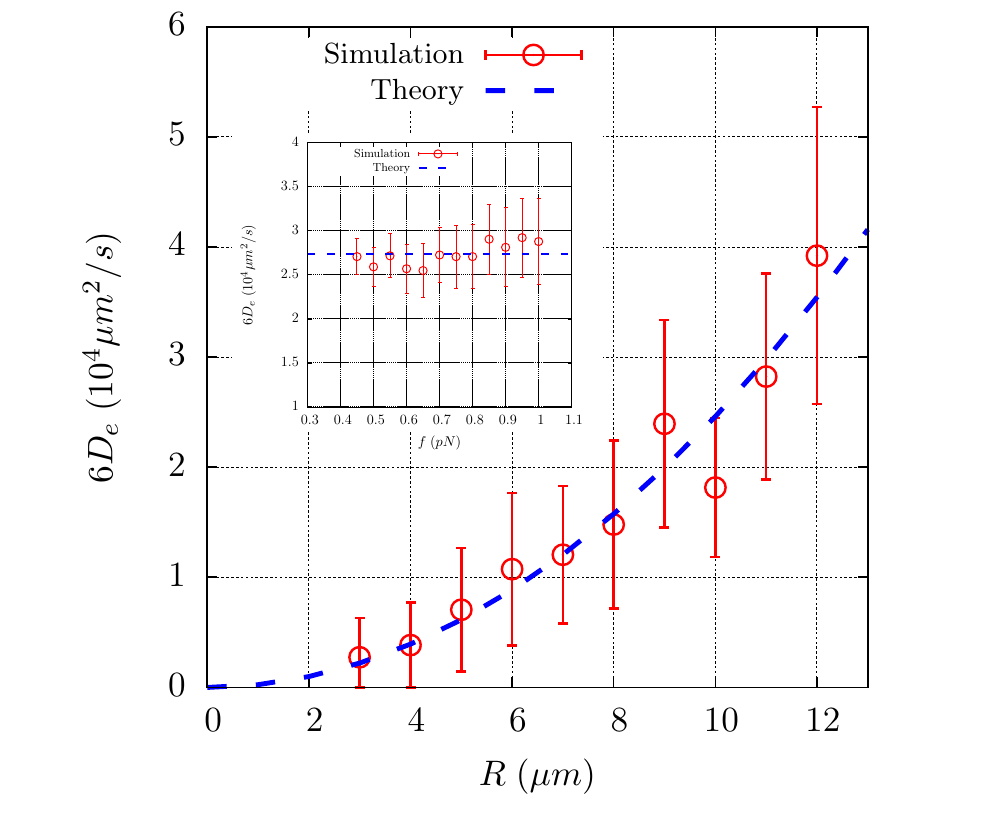}
     \caption{\label{fig:eff_diff_1} Effective diffusion coefficient of bacteria-driven swimmer as a function of the bead radius, $R$, for fixed density of cells $\rho=1/12\,\mu$m$^{-2}$. The parameters $f$, $\lambda_{_T}$ and
     $\lambda_{_R}$ are the same as in Fig.~\ref{fig:helices_2}. Inset shows $D_e$ as a function of the propulsive force $f$, for a micro-swimmer of radius $R=10/,\mu\text{m}$ and the same set of parameters. Mean values are shown in circles, error bars represent one standard deviation above and below the mean and the dashed line is the theoretical prediction of
     Eq.~\eqref{eq:diffusion_2}.}
    \end{figure}
    
    We next validate our analytical results against numerical simulations. In Fig. \ref{fig:msd_log_plot} we show a log-log plot of the MSD as a function of time for a micro-swimmer of radius $R=10\,\mu$m, with density of attached bacteria
    $\rho=1/12~\mu \text{m}^{-2}$, a propulsive force $f=0.48$~pN and transition rates $\lambda_{_T}^{-1}=0.9$~s and $\lambda_{_R}^{-1}=0.1$~s. Time is measured in units of $\tau_r\simeq 68\text{s}$. At very short times
    \mbox{$t\lesssim\tau_r/100\sim\lambda_{_T}^{-1}=0.9\,\text{s}$}, the force and torque remain constant and therefore the MSD is simply the instantaneous mean squared velocity multiplied by the time squared, that is
    \mbox{$\mean{r^2}=3U_e^2t^2$} (dashed magenta line). The solid blue line is Eq.~\eqref{eq:MSD_1} and,  for times-scales shorter than $\tau_r$,   is ballistic   with a MSD 
    approximately  \mbox{$\mean{r^2}\simeq U_e^2t^2$}. On the other hand, for time-scales larger than $\tau_r$, the MSD approaches the asymptotic limit \mbox{$\mean{r^2}\simeq6D_et$} represented by the dashed dotted green line, with
    $D_e$ given by Eq.~\eqref{eq:diffusion_2}. 
    
    On Fig.~\ref{fig:vel_corr} we illustrate the exponential decay of the velocity correlation function (solid red line) for the same micro-swimmer. The dashed line is the theoretical
    prediction from Eq.~\eqref{eq:poly_1} with $D_r=(2\tau_r^{-1})$ and $U_e$ given by Eqs.~\eqref{eq:rot_diff_11} and \eqref{eq:eff_vel_1} respectively. The shaded region is the (numerical) standard deviation. The coarse-grained
    model captures accurately the late time behaviour of the computational results. Finally, Fig.~\ref{fig:eff_diff_1} shows the dependence of $D_e$ with the bead radius $R$, for a micro-swimmer with the same set of parameters $f$,
    $\rho$, $\lambda_{_R}$ and $\lambda_{_T}$. The red circles and error bars represent the numerical mean and the standard deviation of $D_e$, which where obtained from a linear fit to the MSD in the regime $t>3\tau_r$. The dashed
    blue line represents our theoretical prediction Eq.~\eqref{eq:diffusion_2}, which agrees very well with the numerical results. The inset in Fig.~\ref{fig:eff_diff_1} confirms that $D_e$ is independent of the propulsive force of
    the individual bacteria, $f$.
  \subsection{\label{sec:chemotaxis} Chemotaxis of bacteria-driven micro-swimmers}
   In the previous sections, we analysed the motion of the micro-swimmers in chemically-homogeneous environments. Here we consider the case where instead the surrounding environment has a (weak) gradient of a solute to which the cells
   respond chemotactically. The question then emerges, whether the chemotaxis of individual cells translates to chemotaxis of the particle to which they are attached? The answer is not obvious \textit{a priori} since bacteria in
   diametrically opposite positions on the particle can modify its motion in a symmetric fashion and as a result the chemotactic response can be vanishing. Experiments  show however that the
   micro-swimmers perform indeed chemotaxis \cite{zhuang2016, zhuang2017propulsion}. Here we rationalise this result using the coarse grained model developed in the last section and we give an analytical expression for the chemotactic drift, which is validated against numerical simulations.

   \subsubsection{\label{subsect:chem_micro} Chemotactic drift speed}
\begin{figure}
     \centering
     \includegraphics[clip, trim=.6cm 0cm 1.1cm 0cm, width=0.9\columnwidth]{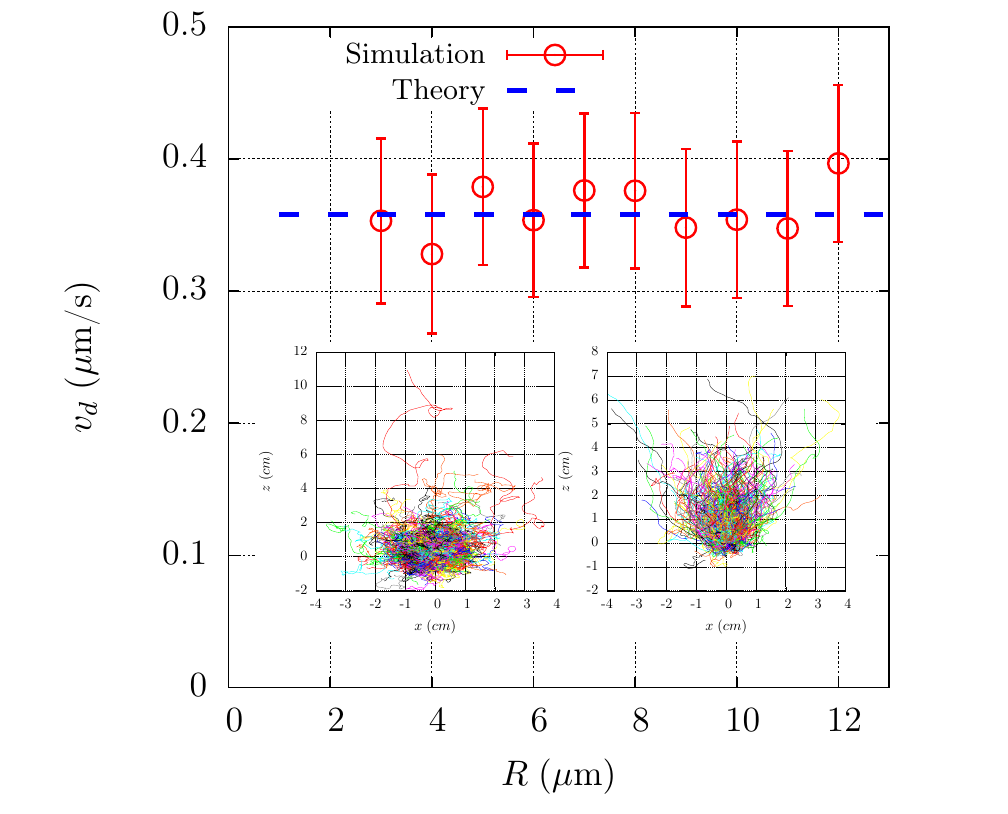}
     \caption{Drift speed of bacteria-driven micro-swimmers as a function of the bead radius, $R$, for fixed cell density. The parameters $\rho$, $f$ and $\lambda_{_R}$ are the same as in Fig.~\ref{fig:helices_2},
     $\lambda_0^{-1}=0.9$~s and $\kappa|\nabla c|=0.005\, \mu$m$^{-1}$. Symbols show mean values from the simulations, error bars represent the standard deviation and the dashed line is the theoretical prediction in 
     Eq.~\eqref{eq:chem_11_1}. The inset shows the biased trajectories in the direction of the chemical gradient for $\kappa|\nabla c|=0.01$ (left) and $\kappa|\nabla c|=0.05$ (right) with each figure showing $10^3$ trajectories
     for a micro-swimmer of radius $R=10\,\mu\text{m}$ and the same set of parameters. In all cases, the simulation time is \mbox{$t=50/(3D_r)\simeq 2240\,\text{s}$}, where $D_r$ is given by Eq.~\eqref{eq:rot_diff_11}.}
     \label{fig:drift_2}
    \end{figure}

    As illustrated in the inset of Fig.~\ref{fig:drift_2}, when the tumbling rate varies according to Eq.~\eqref{eq:chem_1}, the trajectories are biased in the direction of increasing concentration (positive $z$) and the drift increases
    with the magnitude of the chemical gradient. This result can be rationalised observing the dependence of the average displacement on the magnitude of the chemical gradient. Using the coarse-grained model developed in
    Sec.~\ref{subsec:coarse} and following the classical de Gennes analysis for a single cell~\cite{deGennes2004, Locsei2007, Taktikos2013, Pankratova2018}, we will compute the chemotactic drift speed $v_d$. The drift speed is obtained by averaging the mean distance travelled
    in a particular run, over the possible directions that the run can take and dividing the result by the average flight time. As the tumbling rate does not change when the bead moves in the plane perpendicular to $\nabla c$, the average
    displacement is non vanishing only in the direction of the gradient (\textit{i.e.}~$z$). Hence the drift velocity is
    \begin{equation}
     \label{eq:chem_2_1}
     v_d=\frac{\mean{z}}{T_0}=\frac{1}{T_0}\mean{\int_{0}^{\infty}{z(t)Q(t)\text{d}t}}_{\hat{\mathbf{u}}},
    \end{equation}
    \noindent where $Q(t)\text{d}t$ is the probability of tumbling between $t$ and \mbox{$t+\text{d}t$} and the subscript $\hat{\mathbf{u}}$ denotes that averages are taken over the directions of the velocity. Let us denote by $P(t)$ the probability of not stopping
    within $t$ seconds. By the same argument we used to compute $\mathcal{C}(t)$ in Sec.~\ref{subsec:coarse} we know that $P(t)$ is exponentially decaying and decays with rate $1/T$, \textit{i.e.}
    \begin{equation}
     \label{eq:chem_3_1}
     P(t)=\exp{\left[-\int_{0}^{t}{\frac{1}{T(t')}\text{d}t'}\right]},
    \end{equation}
    \noindent and $P(0)=1$. The probability of stopping after $t$ seconds is therefore given by
    \begin{equation}
     \label{eq:chem_4_1}
     1-P(t)=\int_0^t{-\frac{dP(t')}{\text{d}t'}\text{d}t'}=\int_0^t{Q(t')\text{d}t'}.
    \end{equation}
    Substitution of Eqs.~\eqref{eq:chem_3_1} and \eqref{eq:chem_4_1} into Eq.~\eqref{eq:chem_2_1} and using integration by parts gives the result
    \begin{equation}
     \label{eq:chem_5_1}
     v_d=\frac{1}{T_0}\mean{\int_{0}^{\infty}{w(t)\exp{\left[-\int_{0}^{t}{\frac{1}{T(t')}\text{d}t'}\right]}\text{d}t}}_{\hat{\mathbf{u}}},
    \end{equation}
    \noindent where $w=\text{d}z/\text{d}t$ is the velocity of the particle in the direction of the gradient. Remembering that $T^{-1}=3D_r$ from Eq.~\eqref{eq:poly_5}, and using Eq.~\eqref{eq:chem_1} we have
    \begin{equation}
     \label{eq:chem_1_1}
     \frac{1}{T(t)}=\frac{1}{T_0}\left(1-\int_{-\infty}^{t}{K(t-t')c(t')\text{d}t'}\right)+\mathcal{O}\left(|\nabla c|\frac{\lambda_0}{\lambda_{_R}}\right),
    \end{equation}
    \noindent this means that for shallow gradients, the tumbling rate for the micro-swimmer decreases in the same fashion as the tumbling rate for an individual bacterium, as long as $\lambda_{0}\ll\lambda_{_R}$. We use
    Eq.~\eqref{eq:chem_1_1} to expand $P(t)$ up to linear order in the gradient as follows
    \begin{align}
     \label{eq:chem_6_1}
     P(t)&\simeq e^{-t/T_0}+e^{-t/T_0}\frac{|\nabla c|}{T_0}\int_{0}^{t}{\int_{-\infty}^{t'}{K(t'-t'')z(t'')\text{d}t''}\text{d}t'},
    \end{align}
    \noindent where we have ignored the constant $c_0$ which gives terms that do not contribute to the drift, when we consider adaptive kernels with $\int{K(t)\text{d}t}=0$. Inserting Eq.~\eqref{eq:chem_6_1} in Eq.~\eqref{eq:chem_5_1}
    and expressing $K(t)$ as a superposition of impulses we obtain the drift velocity as
    \begin{equation}
     \label{eq:chem_7_1}
     v_d=\int_{0}^{\infty}{\int_{0}^{t}{\int_{0}^{\infty}{\int_{0}^{t'-s}{\frac{|\nabla c|e^{-t/T_0}}{T_0^2}K(s)\mean{w(t)w(s')}_{\hat{\mathbf{u}}}\text{d}s'}\text{d}s}\text{d}t'}\text{d}t}.
    \end{equation} 
    Here we have assumed that the distribution of $\hat{\mathbf{u}}$ is still uniform at first order in $|\nabla c|$, hence $\mean{w}_{\hat{\mathbf{u}}}=0$  and the velocity correlation along $z$ is given by
    \begin{equation}
     \label{eq:chem_8_1}
     \mean{w(t)w(s')}_{\hat{\mathbf{u}}}=\frac{U_e^2}{3}\mean{\hat{\mathbf{u}}(t)\cdot\hat{\mathbf{u}}(s')}_{\hat{\mathbf{u}}}=\frac{U_e^2}{3}e^{-2D_r|t-s'|},
    \end{equation}
    \noindent where we have used the fact that $\hat{\mathbf{u}}$ is spherically symmetric. Substitution of Eq.~\eqref{eq:chem_8_1} into Eq.~\eqref{eq:chem_7_1} yields the drift as
    \begin{align}
     \label{eq:chem_9_1}
     v_d&=\frac{|\nabla c|U_e^2}{6T_0^2D_r}\int_{0}^{\infty}{e^{-\left(1/T_0+2D_r\right)t}\int_{0}^{t}{\int_{0}^{\infty}{K(s)\left(e^{-2D_r(s-t')}-1\right)\text{d}s}\text{d}t'}\text{d}t}\nonumber\\
     &=\frac{|\nabla c|U_e^2}{6T_0^2D_r}\int_{0}^{\infty}{K(s)e^{-2D_rs}\text{d}s}\int_{0}^{\infty}{e^{-\left(1/T_0+2D_r\right)t}\int_{0}^{t}{e^{2D_rt'}\text{d}t'}\text{d}t}\nonumber\\
     &=\frac{|\nabla c|U_e^2}{12T_0^2D_r^2}\int_{0}^{\infty}{K(s)e^{-2D_rs}\text{d}s}\int_{0}^{\infty}{e^{-t/T_0}-e^{-\left(1/T_0+2D_r\right)t}\text{d}t}\nonumber\\
     &=\frac{|\nabla c|U_e^2}{2D_r(3+6D_rT_0)}\int_{0}^{\infty}{K(s)e^{-2D_rs}\text{d}s}.
    \end{align}
        It is known that $K(t)$ vanishes for $t>4 s$~\cite{Segall1986}. Furthermore, $D_r$ decreases quadratically with increasing $R$, therefore for sufficiently large radius  of the particle we have $2D_r s\ll1$ for all relevant values of
    $s$ (for example, when $R=10\,\mu \text{m}$ and $\rho=1/12 \,\mu \text{m}^{-2}$, $2D_r\simeq0.015\,\text{s}^{-1}$) and we may then expand the exponential in the last line of Eq.~\eqref{eq:chem_9_1} to first order in the rotational
    diffusion coefficient. This leads to the general prediction for  drift velocity
    \begin{equation}
     \label{eq:chem_10_1}
     v_d=-\frac{|\nabla c|U_e^2}{3+6D_rT_0}\int_{0}^{\infty}{K(s)s\text{d}s}=-\frac{|\nabla c|U_e^2}{5}\int_{0}^{\infty}{K(s)s\text{d}s}.
    \end{equation}
    \noindent where we have used Eqs.~\eqref{eq:poly_5} which leads to $T_0^{-1}=3D_r$. 
    Our prediction for $v_d$  is seen to be independent of the radius of the particle and is linear in the concentration gradient.

    \begin{figure}
     \centering
     \includegraphics[clip, trim=1cm 0cm 1cm 0.3cm, width=0.9\columnwidth]{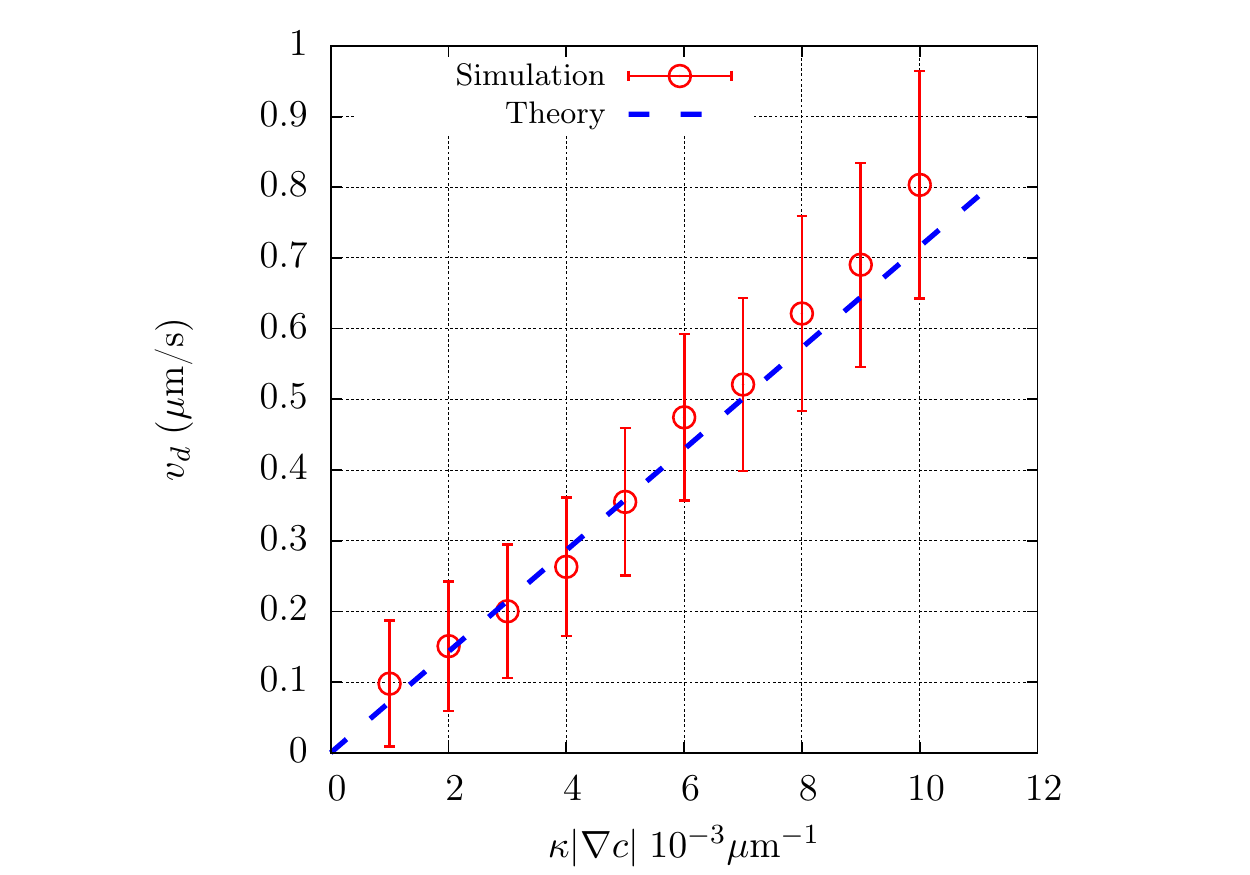}
     \caption{Drift speed of bacteria-driven micro-swimmers as a function of $\kappa|\nabla c|$ for a particle of fixed radius $R=10\,\mu \text{m}$, The parameters $\rho$, $f$ and $\lambda_{_R}$ are the same as in
     Fig.~\ref{fig:helices_2}. The remaining parameters are the same as in Fig.~\ref{fig:drift_2}. Symbols show mean values from the simulations, error bars represent the standard deviation and the dashed line is
     our prediction from Eq.~\eqref{eq:chem_11_1}.}
     \label{fig:drift_3}
    \end{figure}
    For the particular choice of $K(t)=\kappa[\delta(t-t_1)-\delta(t-t_2)]$ used in our computations,  we obtain
    \begin{align}
     \label{eq:chem_11_1}
     v_d=\frac{\kappa|\nabla c|U_e^2(t_2-t_1)}{5}\cdot
    \end{align}
    These results are validated against our  simulations in Figs.~\ref{fig:drift_2} and \ref{fig:drift_3}. In Fig.~\ref{fig:drift_2} we first plot $v_d$ as a function of $R$ for fixed $\kappa|\nabla c|$, \mbox{$t_1=1\,\text{s}$},
    \mbox{$t_2=3\,\text{s}$}, \mbox{$\lambda_0^{-1}=0.9$~s} and the same set of parameters $f$, $\rho$ and $\lambda_{_R}$ as in the isotropic case. Correspondingly, we plot in Fig.~\ref{fig:drift_3} the drift speed as a function of $\kappa|\nabla c|$
    for fixed radius $R=10\,\mu$m and the same set of parameters. Once again, the red circles and error bars represent the numerical values of the mean and standard deviation of $v_d$. These values were obtained from a linear fit to
    the average displacement in the regime $t>10\tau_r$. The dashed blue lines represent the prediction of Eq.~\eqref{eq:chem_11_1} and we can observe that the agreement with the numerical results is excellent.
 \section{\label{sec:discussion} Discussion}
  In this paper, we developed a coarse-grained model that allows   to accurately predict the diffusive behaviour of bacteria-driven micro-swimmers, both in homogeneous environments and in the presence of (weak) chemical gradients. The
  analytical expressions for the effective diffusion coefficient $D_e$ (Eq.~\ref{eq:diffusion_2}) and the chemotactic drift speed $v_d$ (Eq.~\ref{eq:chem_11_1}) were validated against numerical simulations using the stochastic model
  presented in Section \ref{subsec:model} with Eq.~\eqref{eq:chem_1} for the chemotactic change of the tumbling rate of the bacteria. We   found that the effective swimmer velocity, $U_e$, is independent of the  swimmer size, but
  increases with the square root of the number of surface-attached bacteria $N$, when $N$ is large. On the other hand, the rotational diffusion coefficient, $D_r$, increases linearly with $N$ and is inversely proportional to the particle
  radius $R$, in contrast with the thermal rotational diffusion coefficient $D_{r}^T$ which decreases as $R^{-3}$. For shallow chemical gradients $\kappa|\nabla c|R\ll1$, we found that the micro-swimmers respond chemotactically if we
  assume that each bacterium on its surface does, and that the resulting chemotactic drift speed is proportional to $|\nabla c|$ and independent of $R$, the latter being a consequence of the bacteria being constrained by the bead and
  each bacterium experiencing the same change in concentration.
 
  The expression  we obtained for the effective velocity in this paper can be directly compared with experimental results. Behkam et al.~reported $U_e=14.8\pm 1.1\,\mu \text{m/s}$ for micro-swimmers of radius $R=10\,\mu\text{m}$ and
  densities of attachment between one bacterium per $12\,\mu\text{m}^2$ and $7\,\mu\text{m}^2$ with $f\simeq0.48\,\text{pN}$ \cite{Behkam2008}. In comparison, our model predicts $U_e\simeq 12.99-17.42\, \mu\text{m/s}$, in good
  agreement. The same study also reported that the swimming speed increases with the square root of the number of bacteria $N$, for $N\geq10$, as predicted theoretically.   The same scaling result was reported by Arabagi et
  al.~\cite{arabagi2011modeling} and Zhuang et al.~\cite{zhuang2017propulsion}. Furthermore, both of these studies reported swimming velocities independent of the particle radius in homogeneous environments as well as in the presence
  of a chemical gradient, again confirming our theoretical results. Unfortunately there are no past experimental measurements of the effective diffusion coefficient. In their numerical simulations however, Arabagi et al.~\cite{arabagi2011modeling}
  reported the value $D_e=15185\,\mu\text{m}^2/\text{s}$ for a bead of radius $R=20\,\mu\text{m}$ and $f\simeq 0.48\,\text{pN}$,  in good agreement with Eq.~\eqref{eq:diffusion_2} which predicts $D_e\simeq15802\,\mu\text{m}^2/\text{s}$.
  Regarding the chemotactic drift of particles with surface-attached bacteria, Zhuang et al.~\cite{zhuang2016, zhuang2017propulsion} found that $v_d$ is proportional to the magnitude of the chemical gradient and that it is independent of
  the particle size, both of which are predicted by Eq.~\eqref{eq:chem_11_1}.
 
  The analytical results in this paper will be useful for the practical design of bacteria-driven micro-swimmers. The model can be extended to include other steering strategies such as phototaxis, pH-taxis and external electromagnetic
  fields~\cite{Steager2007, Steager2011}. 
  The assumption of a chemically permeable  particle could also be relaxed, and the perturbation to the concentration field by the presence of the particle (and the cells) should be solved for.
  Another possible extension is the study of deformable random walkers, inspired by experiments with bacteria-driven micro-swimmers in which \textit{E.~coli} are attached to red blood cells
  \cite{Alapan2018}. One of the ultimate applications of bacteria-driven micro-swimmers is targeted drug delivery, which requires a suspension of these particles to move collectively. A mathematical description of the collective
  behaviour of multiple micro-swimmers is therefore needed and this requires an explicit consideration of their hydrodynamic interactions. The interactions depend critically on the distribution of attached bacteria on the surface of
  each particle, which can be controlled using micro-manipulation \cite{Barroso2015} and nanoprinting methods~\cite{Carlsen2014, Rozhok2005}. A theory including hydrodynamic interactions, taking clues from classical work on active fluids
  \cite{marchetti_review,julicher2018hydrodynamic} will therefore allow to move bacteria-driven micro-swimmers closer to applications.

 \section*{Conflicts of interest}
  There are no conflicts to declare.
 
 \section*{Acknowledgements}
  This project has received funding from the European Research Council (ERC) under the European Union's Horizon 2020 research and innovation programme  (grant agreement 682754 to EL).
 \appendix
 \section{\label{appx:alpha_0} Singular case $\alpha_i=0$}
  When all bacteria point in the radial direction, there is no applied torque and every vector $\mathbf{p}_i$ remains constant in time. In this case, the velocity correlation function is easily computed as follows
  \begin{align}
   \label{eq:alpha_0_1}
   \mean{\mathbf{V}(0)\cdot\mathbf{V}(t)}&=\left(\frac{f}{6\pi\mu R}\right)^2\sum_{i,j}{\mean{\epsilon_i(0)\epsilon_j(t)}\mathbf{p}_i\cdot\mathbf{p}_j}  \nonumber\\
   &=N\left(\frac{f}{6\pi\mu R}\right)^2p_{_R}p_{_{RR}}(t),
  \end{align}
  where we have used the fact that the orientations and positions of bacteria are uncorrelated. Integrating twice, we obtain the MSD
  \begin{align}
   \label{eq:alpha_0_2}
   \mean{r^2(t)}&=2N\left(\frac{fp_{_R}^s}{6\pi\mu R}\right)^2\frac{t^2}{2}\nonumber\\
   &+2N\left(\frac{fp_{_R}^s}{6\pi\mu R}\right)^2\int_0^t{\int_{s_1}^t{\left(\frac{\lambda_{_T}}{\lambda_{_R}}e^{-(\lambda_{_R}+\lambda_{_T})(s_2-s_1)}\right)\text{d}s_2}\text{d}s_1}\nonumber\\
   &=2N\left(\frac{fp_{_R}^s}{6\pi\mu R}\right)^2\frac{t^2}{2}\nonumber\\
   &+2N\left(\frac{fp_{_R}^s}{6\pi\mu R}\right)^2\frac{\lambda_{_T}}{\lambda_{_R}\left(\lambda_{_T}+\lambda_{_R}\right)}\left(t+\frac{e^{-(\lambda_{_R}+\lambda_{_T})t}-1}{\left(\lambda_{_T}+\lambda_{_R}\right)}\right).
  \end{align}
  This is essentially ballistic and for $1\ll t\left(\lambda_{_T}+\lambda_{_R}\right)$ we have $\mean{r^2(t)}\simeq 3U_e^2 t^2\equiv L^2(t)$ with $U_e$ given by Eq.~\eqref{eq:eff_vel_1}. 
  
  This result might appear counter
  intuitive, as one would expect that a large number of uniformly distributed radial forces acting on a sphere should account for no net thrust. However, this is only true if the time autocorrelation of the forces decays to zero.
  As the autocorrelation of the epsilon variables decays to a finite value, $(p_{_R}^s)^2$ from Eq.~\eqref{eq:prob_6}, so does the velocity correlation function, hence the ballistic motion. Tilting the forces away from the
  radial direction allows for reorientation of the sphere, reducing the probability of finding a given force pointing in the same direction at two different times accounting for a vanishing autocorrelation at large time-scales, hence
  the long-time diffusive behaviour for the case $\alpha_i\neq0$. Notice that $L(t)$ is the maximum average length that a trajectory can have (with $\alpha_i$ not necessarily null). Indeed, the arc length of a helix is
  the same as that of a straight line traversed at the same speed, therefore $L(t)$ can be obtained by considering the square root of Eq.~\eqref{eq:eff_vel_0} in the limit when $\mean{\mathbf{V}}_{\epsilon}$ is parallel to
  $\mean{\boldsymbol{\omega}}_{\epsilon}$ and multiplying the result by $t$, which is exactly $L(t)=\mean{|\mathbf{V}_{\epsilon}|^2}^{1/2}t=\sqrt{3}U_et$.

\balance

 \bibliography{bacteria_driven_microswimmers_1.bib} 
\bibliographystyle{rsc} 

\end{document}